\documentclass[prb,preprint, footinbib, longbibliography]{revtex4-1}  
\usepackage{setspace}
\usepackage{xcolor}
\usepackage{soul}
\sethlcolor{yellow}
\usepackage[english]{babel}
\usepackage{epsf,color,colordvi}
\usepackage{graphicx}
\usepackage{amsmath}
\usepackage{amssymb}
\usepackage{times}
\usepackage{xcolor}
\usepackage[normalem]{ulem}
\usepackage{comment}

\usepackage{subfigure}
\usepackage{placeins}

\newcommand{\ts}{\textstyle }
\newcommand{\be}[1]{\begin{equation}\label{eq:#1}}
\newcommand{\cH}{{\cal H}}
\newcommand{\cR}{{\cal R}}
\newcommand{\cI}{{\cal I}}
\newcommand{\cP}{{\cal P}}
\newcommand{\ee}{\end{equation}}
\newcommand{\mr}[1]{\mathrm{#1} }
\newcommand{\ds}{\displaystyle}

\newcommand{\SE}{Schr\"{o}dinger equation}

\newcommand{\beq}{\begin{equation}}
\newcommand{\eeq}{\end{equation}}
\newcommand{\br}{\textbf{r}}

\newcommand{\cO}{{\cal O}}
\newcommand{\cM}{{\cal M}}
\newcommand{\cL}{{\cal L}}
\renewcommand{\i}{$i$}
\renewcommand{\br}{\textbf{r}}

\newcommand{\dt}{\Delta t}

\begin{document}

\title{On numerical solutions of the time-dependent  Schr\"odinger equation }

\begin{abstract}
We review an explicit approach to obtaining numerical solutions of the \SE~that is conceptionally straightforward and capable of significant accuracy and efficiency.  The method and its efficacy are illustrated with several examples.  Because of its explicit nature, the algorithm can be readily extended to systems with a higher number of spatial dimensions. We show that the method also generalizes the staggered-time approach of Visscher and allows for the accurate calculation of the real and imaginary parts of the wave function separately.
\end{abstract}
\mbox{}\\

\author{Wytse van Dijk}
\email{vandijk@physics.mcmaster.ca}
\affiliation{Department of Physics and Astronomy, McMaster University, Hamilton, Ontario, Canada L8S 4M1}
\affiliation{Redeemer University, Ancaster, Ontario, Canada L9K 1J4}

\maketitle

\section{Introduction}\label{sec:01}

The time-dependent \SE\ can be written as
\beq
\label{1.01}
\left(i\hbar\dfrac{\partial~}{\partial t}-H\right)\psi(\br,t) = 0,
\eeq
with
\beq
\label{1.01b}
\psi(\br,t_0)=\phi(\br).
\eeq
$H(\textbf{r})$ is the time-independent Hamiltonian operator
\beq
\label{1.02}
H = -\dfrac{\hbar^2}{2m}\nabla^2 + U(\br),
\eeq
$\hbar$ is the Planck constant, $m$ is the mass of the particle, and $U(\br)$ is the potential energy. The solution of Eq.~\eqref{1.01} has the initial wave function at time $t_0$ as input.  We can show that 
\beq
\label{1.04} 
\psi(\br,t+\Delta t)  = e^{\ts -iH\Delta t/\hbar}\psi(\br,t)
\eeq
satisfies Eq.~\eqref{1.01} for any $\Delta t$.
The solutions of Eq.~\eqref{1.01}  describe a large variety of nonrelativistic atomic and subatomic systems.    Usually a limited number of closed-form solutions are discussed, but these are few and often describe approximations of   physical systems.  Although these solutions provide valuable insight,  we need to solve the \SE \ numerically to determine accurate wave functions or to make predictions of experimentally accessible data.

A variety of numerical methods for treating the time behavior are based on approximations of the time-evolution operator $e^{\ts -iH\Delta t/\hbar}$.  The simplest way  to proceed is to consider the first-order Taylor expansion of this operator, which leads to the forward Euler method,
\beq
\label{1.05} 
\psi(\br,t+\Delta t) = (1-iH\Delta t/\hbar)\psi(\br,t) + \cO[(\Delta t)^2]. 
\eeq
A  successful and accurate approximation is one by Tal-Ezer and Kosloff~\cite{talezer84} which expands the time-evolution operator in terms of Chebyshev polynomials.

Another effective approach is the exponential split-operator approach (see, for example, Ref.~\onlinecite{neuhauser09}) in which the kinetic and potential energy operators in the exponent are separated.  This separation  becomes cumbersome for higher orders because the operators do not commute.  The Crank-Nicolson method~\cite{crank47} and its generalization~\cite{vandijk07} preserve normalization, but the method is implicit, which   makes it computationally less efficient.  Although explicit methods express the later-time wave function in terms of an operator acting on the earlier time one(s), implicit methods 
require linear systems of equations involving the earlier and later wave functions to be solved.

The simple forward Euler method in Eq.~\eqref{1.05} has stability issues.  To overcome this  problem  Askar and Cakmak~\cite{askar78}  suggested calculating the wave function in terms of the wave functions at two earlier times by combining Eq.~\eqref{1.04} with
\beq
\label{1.06}
\psi(\br,t-\Delta t)=e^{\ts iH\Delta t/\hbar}\psi(\br,t)   
\eeq
to obtain
\begin{subequations}
\label{1.07}
\begin{align}
\psi(\br,t+\Delta t) & = \psi(\br,t-\Delta t) 
-\left(e^{\ts iH\Delta t/\hbar}-e^{\ts -iH\Delta t/\hbar}\right)\psi(\br,t) \label{1.07a} \\
& = \psi(\br,t-\Delta t) - 2i\dfrac{\Delta tH}{\hbar} \psi(\br,t)+\cO[(\Delta t)^3]. \label{1.07b}
\end{align}
\end{subequations}
The possibility of stability is ensured,~\cite{askar78,rubin79} and as an added benefit, the order of the error has gone from $\cO[(\Delta t)^2]$ to $\cO[(\Delta t)^3]$, as can be seen by expanding the exponentials in Eq.~(\ref{1.07a})  in a Taylor series. An important, and potentially useful property of the solution is that the normalization and energy integrals are also approximately conserved,
\beq
\label{1.08}
\int \psi^*(\br,t) {\cal A}\psi(\br,t)\; dV = \mbox{constant},
\eeq
where ${\cal A} = I$ or $H$ and $I$  is the identity operator.   

The goal of this paper is to discuss an expansion of the time-evolution operator that leads to a simple explicit method with potentially significant and systematic improvements.  We introduce the expansion of the time-evolution operator in Sec.~\ref{sec:02}, consider the spatial integration in Sec.~\ref{sec:03},  discuss some properties of the algorithm in Sec.~\ref{sec:04}, and illustrate the approach by   sample calculations in Sec.~\ref{sec:05}. We introduce some further implications of the approach in Sec.~\ref{sec:06} and provide a conclusion  in Sec.~\ref{sec:disc}.  Several problems are posed in Sec.~\ref{sec:problem}.

\section{Expansion of the time-evolution operator}\label{sec:02}

Because the explicit method of Askar and Cakmak is efficacious, we use it as a starting point to obtain a higher-order expansion in $\Delta t$ for the time-evolution operator as was done in Ref.~\onlinecite{vandijk22}.   We will see that a modification of Eq.~\eqref{1.07} generates highly accurate solutions as functions of time.  We assume in this section that we have a numerical method to be introduced later to calculate $H\psi(\br,t)$ accurately for a given $\br$ and $t$.
 We discretize time as $t=t_n$ for $n=0,1,\dots,N$, where $t_n= n\Delta t$, and determine the wave functions $\psi^n(\br)=\psi(\br,t_n)$.   We write  Eq.~\eqref{1.07} as
\beq
\label{2.01}
\psi(\br,t+\Delta t)=\psi(\br,t-\Delta t)-2i\sin(H\Delta t)\psi(\br,t),
\eeq
so that for $t=t_n$ we have
\beq
\label{2.02}
\psi^{n+1} = \psi^{n-1} -2i\sin(H\Delta t/\hbar)\:\psi^n.
\eeq
 A Taylor series expansion yields a polynomial approximation of the sine function, which we write as
\begin{subequations}
 \label{2.03}
\begin{align}
\sin(z) & = z\left(1-\dfrac{1}{3!}z^2 + \dfrac{1}{5!}z^4
 - \cdots + \dfrac{(-1)^M}{(2M+1)!}z^{2M}\right)+  ~\cO(z^{2M+3}) \\
& \ds = z\prod_{s=1}^{2M}\left(1-\dfrac{z}{z^{(2M)}_s}\right) + \cO(z^{2M+3}),
\end{align}
\end{subequations}
where  $z^{(2M)}_1,z^{(2M)}_2,\dots,z^{(2M)}_{2M}$ are the $2M$ zeros of the $2M$th-order polynomial approximation of $\sin z/z$. 
Symbolic mathematics software such as MAPLE calculates the numerical values of the zeros to the desired precision.  Several  $M$-value zeros are shown in Table~\ref{table:01a}.

\begin{table}[ht]
\begin{tabular}{c|cccccc}
\hline\hline
$M$ & $s=1$ & 2 & 3 & 4 & 5 & 6 \\
\hline
1 & $-2.44949+i0.00000$ & $2.44949+i0.00000$ \\
2 & $3.23685-i0.69082$ & -3.23685+\i0.69082 & $3.23685+i0.69082$ & $-3.23685-i0.69082$ \\
3 & ~~$3.07864+i0.00000$~~ & ~~-$3.07864+i0.00000$~~ & ~~$4.43401-i1.84375$~~ & ~~$-4.43401+i1.84375$~~ & ~~$4.43401+i1.84375$~~ & ~~$-4.43401
-i1.84375$~~\\
\hline\hline
\end{tabular}
\caption{The zeros $z_s^{(2M)}$, $s=1,2,\ldots,2M$ of the polynomial approximation of $\sin z/z$ for $M=1,2,3$. \label{table:01a}}
\end{table}
 
If we define
\beq
\label{2.05}
K^{(2M)}_s \equiv 1-\dfrac{H\Delta t/\hbar}{z^{(2M)}_s}, \ \ s=1,2,\dots,2M,
\eeq
 Eq.~\eqref{2.02} becomes
\beq
\label{2.06}
\psi^{n+1} = \psi^{n-1} - 2iH\Delta t/\hbar\prod_{s=1}^{2M}K_s^{(2M)}\psi^n.
\eeq
It is convenient to employ a recursive procedure  for $s=1,2,\dots,2M$,
\beq
\label{2.07}
\psi^{(n,s)} =K_s^{(2M)}\psi^{(n,s-1)}   \mbox{ with } \psi^{(n,0)}=\psi^n, 
\eeq
so that
\beq
\label{2.08}
\psi^{n+1}=\psi^{n-1}-(2iH\Delta t/\hbar)\psi^{(n,2M)}.
\eeq
Despite the recursion, the method remains explicit.

\begin{figure}[htpb]
  \centering
   \resizebox{3.5in}{!}{\includegraphics[angle=-90]{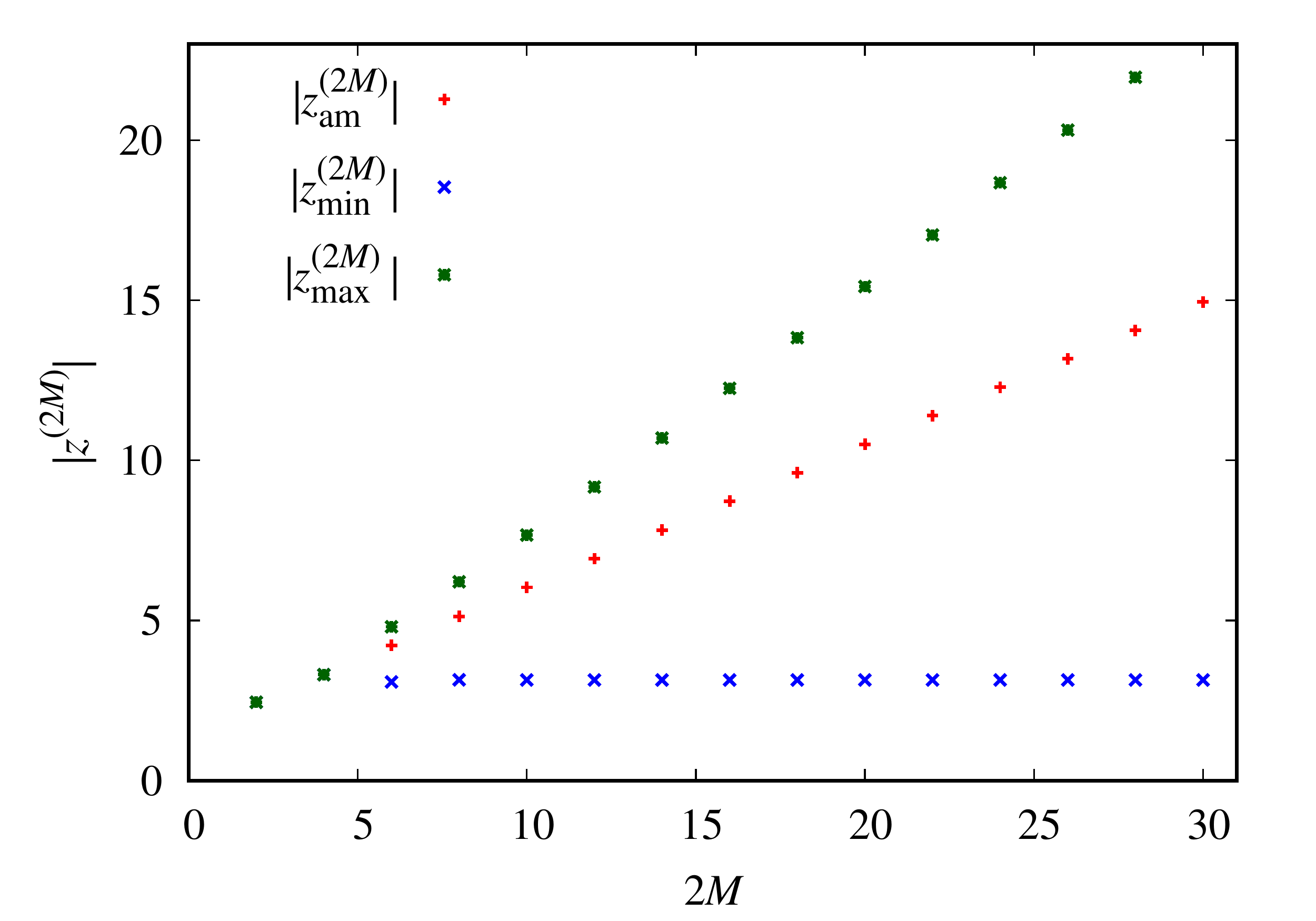}}
\caption{The arithmetic mean, minimum, and maximum values of $|z^{(2M)}_s|$ for each $2M$.}
\label{fig:01a}
\end{figure}

To estimate the effect of larger values of $M$ on the precision and the efficiency of the expansion, we plot in  Fig.~\ref{fig:01a}  the arithmetic mean
\beq
\ds|z_\mr{am}^{(2M)}| = \left(\dfrac{1}{2M}\sum_{s=1}^{2M} |z_s^{(2M)}|\right),
\eeq
the minimum $|z_{\min}^{(2M)}|$, and the maximum $|z_{\max}^{(2M)}|$ of the set $\{|z_1^{(2M)}|,|z_2^{(2M)}|,\dots,|z_{2M}^{(2M)}| \}$  as functions of $M$.  Because $K_s^{(2M)}$ depends on $\Delta t/z_s^{(2M)}$ rather than $\Delta t$, we see from Fig.~\ref{fig:01a}  that 
$\Delta t/z_s^{(2M)}$  decreases with $M$, and thus including more terms provides a more precise expansion, which allows for larger values of $\Delta t$.

\section{Spatial integration}\label{sec:03}

To implement the time advance algorithm, Eq.~(\ref{2.08}), we need to numerically evaluate $H\psi$, where the operator $H$ involves second-order spatial derivatives.  A variety of methods can be employed, including a fast Fourier spectral approach,\cite{vandijk11} high-order-compact differencing,\cite{tian10} and differencing on a nonuniform grid.\cite{lynch05}  A simple and efficient method is the central differencing approach, which is also useful for studying the stability of the algorithm. 

For now we focus on one spatial dimension.     
We begin by discretizing the spatial and temporal variable of the wave function so that $\psi^n_j=\psi(x_j,t_n)$, where  $x_j= x_0+j\Delta x$ for $j=0,1,\dots,J$.  The $\psi^n_j$ can be considered as the components of a 
$J$-dimensional vector representing the discretized $\psi^n$.  Because the Hamiltonian involves the second-order spatial derivative, we approximate the second derivative to order $2r-1$ of $\Delta x$ by a sum of terms from $-r$ to $r$:
\beq
\label{3.01}
\dfrac{\partial^2\psi^n_j}{\partial x^2} = \dfrac{1}{(\Delta x)^2}\sum_{\ell=-r}^r c_\ell^{(r)} \psi_{j+\ell}^n + \cO{[(\Delta x)^{2r}]},
\eeq
where the constants $c_\ell^{(r)}$ are rational numbers independent of $\Delta x$.  For example,  the familiar lowest order approximation ($r=1$) is 

\beq
\dfrac{\partial^2\psi^n_j}{\partial x^2} = \dfrac{1}{(\Delta x)^2}(\psi^{n}_{j+1}+ \psi^{n}_{j-1} - 2\psi^{n}_j), \label{lowest}
\eeq
 corresponding to $c_1^{(1)}= c_{-1}^{(1)} = 1$ and $c_0^{(1)} = -2$.   Because the second derivative depends on $(\Delta x)^2$, we must have $c_{\ell}^{(r)} = c_{-\ell}^{(r)}$.  The  $c_{\ell}^{(r)}$ are obtained by considering
\begin{subequations}
\label{3.01a}
\begin{align}
f^{(2)}(x_j) & = \ds\dfrac{1}{(\Delta x)^2}\sum_{\ell=-r}^{r} c_\ell^{(r)}f(x_j+\ell\Delta x) \\
& =  \ds\dfrac{1}{(\Delta x)^2}\sum_{\ell=-r}^{r} c_\ell^{(r)} \sum_{i=0}^\infty\dfrac{\ell^i(\Delta x)^i}{i!}f^{(i)}(x_j) \\
& =  \ds\dfrac{1}{(\Delta x)^2}\sum_{i=0}^\infty \dfrac{(\Delta x)^i}{i!}f^{(i)}(x_j)\sum_{\ell=-r}^{r} c_\ell^{(r)}\ell^i,  \label{3.01c}
\end{align}
\end{subequations}
 where $f^{(i)}(x)$ is the $i$th derivative of $f(x)$.  Equation~(\ref{3.01c}) consists of a linearly independent combination of the functions $f^{(i)}(x), i=0,1,\dots$, so that $\sum_{\ell=-r}^{r} c_\ell^{(r)}\ell^i = 0$ for $ i \neq 2$ and equals 2 for $i = 2$.  Because $c_{-\ell}^{(r)} = c_\ell^{(r)}$, we obtain
\beq
\label{3.02} 
-\dfrac{c_0^{(r)}}{2}\delta_{0i}+\delta_{2i}=\sum_{\ell=1}^r\ell^{i}c_\ell^{(r)}  \mbox{ for } \qquad (i=0,1,\dots,r).
\eeq
By setting $i = 0$, we obtain
\beq
\label{3.03}
c^{(r)}_0=-2\sum_{\ell=1}^r c_\ell^{(r)}.
\eeq
We let $r=1$ in Eqs~(\ref{3.02}) and (\ref{3.03}) and obtain the lowest order expansion in Eq.~(\ref{lowest}). 
It is simple to generate the coefficients using a symbolic algebra package. The first seven sets of coefficients  are
given in Table~\ref{table1}.

\begin{table}[h]
\begin{center}
\caption{The coefficients $c_\ell^{(r)}$ up to $r=7$.}
\label{table1}
\begin{tabular}{c|cccccccc}
\hline\hline
$r$ & \multicolumn{1}{c}{$\ell=0$~~~} & \multicolumn{1}{c}{$1$} & 
\multicolumn{1}{c}{$2$} & \multicolumn{1}{c}{$3$} & 
\multicolumn{1}{c}{$4$} & \multicolumn{1}{c}{$5$} & 
\multicolumn{1}{c}{$6$} & \multicolumn{1}{c}{$7$}  \\
\hline
1 & $-2$ & 1 \\
2 & $-\frac{5}{2}$ & $\frac{4}{3}$ & $-\frac{1}{12}$ \\
3 & $-\frac{49}{18}$ & $\frac{3}{2}$ & $-\frac{3}{20}$ & $\frac{1}{90}$ \\
4 & $-\frac{205}{72}$ & $\frac{8}{5}$ & $-\frac{1}{5}$ & $\frac{8}{315}$ 
  & $-\frac{1}{560}$ \\
5 & $-\frac{5269}{1800}$ & $\frac{5}{3}$ & $-\frac{5}{21}$ & $\frac{5}{126}$ 
  & $-\frac{5}{1008}$ & $\frac{1}{3150}$ \\
6 & $-\frac{5369}{1800}$ & $\frac{12}{7}$ & $-\frac{15}{56}$ 
  & $\frac{10}{189}$ & $-\frac{1}{112}$ & $\frac{2}{1925}$ 
  & $-\frac{1}{16632}$ \\
~7~ & $-\frac{266681}{88200}$ & $\frac{7}{4}$ & $-\frac{7}{24}$ 
  & $\frac{7}{108}$ & $-\frac{7}{528}$ & $\frac{7}{3300}$ 
  & $-\frac{7}{30888}$ & $\frac{1}{84084}$ \\
\hline
\end{tabular}
\end{center}
\end{table} 

We eventually  need to evaluate $H\psi^n$, which can be written as
\begin{subequations}
\label{3.04}
\begin{align}
H\psi^n_j &= -\dfrac{\hbar^2}{2m}\left(\dfrac{\partial^2\psi^n_j}{\partial x^2}\right) + V_j\psi^n_j \\
& = \ds -\dfrac{\hbar^2}{2m(\Delta x)^2}\sum_{\ell=-r}^r c_\ell^{(r)} \psi^n_{j+\ell} + V_j\psi^n_j.
\end{align}
\end{subequations}
If we express
$H\psi^n = A^{(r)}\psi^n$, 
$A^{(r)}$ is a symmetric and  banded matrix  with $r$ super- and $r$ sub-diagonals, 
\beq
\label{3.05}{\small
A = \left(\begin{array}{cccccccccc}
        d_0 & a_1 & a_2 & \cdots & a_r   & 0 \\
        a_{-1} & d_1 & a_1 &  \cdots & a_{r-1} & a_r \\
        a_{-2} & a_{-1} & d_2 & \cdots & a_{r-2} & a_{r-1} \\
        \vdots & \vdots & \vdots && \vdots & \vdots \\
        a_{-r} & a_{-r+1} & a_{-r+2} & \cdots & d_r & a_1 \\
        0   & a_{-r} & a_{-r+1} & \cdots & a_{-1} & d_{r+1} \\
            &     &     &     &  && \ddots \\
            &     &     &     &  &&&       & d_{J-1} & a_1 \\
            &     &     &     &  &&&       & a_{-1}     & d_J
        \end{array} \right).
}
\end{equation}
The superscript $\mbox{}^{(r)}$ of $A$ and the $a_\ell$ and the $d_j$  have been omitted.    The matrix elements are defined as
\begin{subequations}
\label{305}
\begin{align}
a^{(r)}_\ell & = - \dfrac{\hbar^2}{2m(\Delta x)^2}c_\ell^{(r)}  \\
d^{(r)}_j & = - \dfrac{\hbar^2}{2m(\Delta x)^2} c_0^{(r)} + V_j,
\end{align}
\end{subequations}
where $V_j=V(x_j)$. Thus $\ds K_s^{(2M)} = 1 - \Delta t H/\hbar/z_s^{(2M)}$ is a  banded matrix like $A^{(r)}$, but  is not Hermitian because of the complex $z^{(2M)}_s$. Because of the properties of the product of commuting symmetric banded matrices (see Appendix~\ref{app:02}), the matrix
\beq
\label{306}
B^{(2M,r)}= 2iH\Delta t/\hbar\prod_{s=0}^{2M}K_s^{(2M)}
\eeq
is symmetrically banded with $[(2M+1)r]$ sub- and super-diagonals. For a particular choice of $M$ and $r$, we evaluate the matrix $B^{(2M,r)}$ and generate the discretized wave functions at time $t_n$, using Eq.~\eqref{2.06}
\beq
\label{308}
\psi^{n+1} = \psi^{n-1} - B^{(2M,r)}\psi^n,
\eeq
with $J$-dimensional vectors $\psi^0$ and $\psi^1$ provided as input.

An accurate single-step approach to obtaining $\psi^1$ from $\psi^0$, which we use in the examples is described in  Appendix~\ref{app:01}. 
Note that the matrix $B^{(2M,r)}$ is independent of $t_n$, and is banded so that the number of computations to obtain $\psi$ at each time step is proportional to $[(2M+1)(2r+1)J]$.
We denote the method that uses the matrix $B^{(2M,r)}$ as method~I.

Alternatively, and preferably for  applications, Eq.~\eqref{2.06} is evaluated recursively using Eqs.~(\ref{2.07}) and (\ref{2.08}).  For computational efficiency it is advisable to exploit the fact that the matrices are  banded.  This approach is denoted as method~II.
The generalization to two- and three-dimensional systems is straightforward.\cite{vandijk22}

\subsection{Implementation}

An example of implementing the recursive approach is shown in the  following pseudocode.
In the procedure \textit{$\psi$\_sine\_step} we use complex arrays $(\psi^0_j)$ and $(\psi^1_j)$ to represent the numerical wave functions at $t=n\Delta t$ and $t=(n+1)\Delta t$  to obtain $(\psi^2_j)$, the numerical wave function at $t=(n+2)\Delta t$. 
 
\begin{singlespace}
{\small
\noindent \textbf{procedure} \textit{$\psi$\_sine\_step($\Delta t,M,r,J,(\psi^0),(\psi^1),(\psi^2))$} \\
\textbf{integer} $M, r, J, s$;~~~~\textbf{real} $\Delta t$ \\
\textbf{complex array} $\psi^0_{0:J}, \psi^1_{0:J}, \psi^2_{0:J}, \psi_{0:J},  \overline{h\psi}_{0:J}$   \\
\mbox{~~~}$\psi \leftarrow \psi^1$  \\
\mbox{~~~}\textbf{for} $s = 1$ \textbf{to} $2M$ \textbf{do} \\
\mbox{~~~~~~}\textbf{call procedure} $\overline{H\psi}(J,r,\psi,\overline{h\psi})$  \\
\mbox{~~~~~~} $\psi \leftarrow \psi-\frac{\Delta t}{\hbar}\frac{1}{z_s^{(2M)}}\overline{h\psi}$ \\
\mbox{~~~}\textbf{end for}  \\
\mbox{~~~}\textbf{call procedure} $\overline{H\psi}(J,r,\psi,\overline{h\psi})$  \\
\mbox{~~~} $\psi^2 \leftarrow \psi^0 - 2i \frac{\Delta t}{\hbar}\overline{h\psi}$  \\
\textbf{end procedure} \textit{$\psi$\_sine\_step}	
}
\end{singlespace}

\noindent The variables $z_s^{(2M)}, c_{\ell}^{(r)}, \hbar, m$ and $V(x_j)$ are assumed to be defined and accessible in the procedures.  
The  subprocedure $\overline{H\psi}$ is
\begin{singlespace}
{\small
\noindent \textbf{procedure} $\overline{H\psi}(J,r,\psi,\overline{h\psi})$ \\
\textbf{integer} $J, r, k$;  ~~~\textbf{complex array} $\psi_{0:J}, \overline{h\psi}_{0:J}, \phi_{0:J}$   \\
\mbox{~~~}$\phi\leftarrow 0$  \\
\mbox{~~~}\textbf{for} $j=0$ \textbf{to} $J$ \textbf{do} \\
\mbox{~~~~~~}\textbf{for} $\ell=-r$ \textbf{to} $r$ \textbf{do} \\
\mbox{~~~~~~~~~} \textbf{if} $(j+\ell>=0$ \textbf{AND} $j+\ell=<J)$ \textbf{then}  \\
\mbox{~~~~~~~~~~~~} $\phi_j\leftarrow$ $\phi_j - c_\ell^{(r)}\psi_{j+\ell}$ \\
\mbox{~~~~~~~~~}\textbf{end if}  \\
\mbox{~~~~~~}\textbf{end for}  \\
\mbox{~~~~~~}$\overline{h\psi}_j \leftarrow -\frac{\hbar^2}{2m(\Delta x)^2}\phi_j + V(x_j)\psi_j$  \\
\mbox{~~~}\textbf{end for}  \\
\textbf{end procedure} $\overline{H\psi}$
}
\end{singlespace}

Consider the free traveling wave packet described in Ref.~\onlinecite{vandijk14}:
\beq
\label{2.01a}
\psi(x,\tau) = \sqrt{\dfrac{\alpha}{\sqrt{\pi}}}\dfrac{1}{\sqrt{1+2i\alpha^2\tau}}\exp\left(\dfrac{-\frac{1}{2} \alpha^2x^2+ikx-ik^2\tau}{1+2i\alpha^2\tau}\right),
\eeq
with $\tau = \hbar t/m$.
Analytical calculations show that 
\beq
\label{2.02a}
\langle x\rangle = 2k\tau, \quad \langle x^2\rangle = 4k^2\tau^2 + \dfrac{1+4\alpha^4\tau^2}{2\alpha^2},
\eeq
so that
\beq
\label{2.03a}
(\Delta x)^2 = \langle x^2\rangle - \langle x\rangle^2 
= \dfrac{1+4\alpha^4\tau^2}{2\alpha^2}.
\eeq
Because $\Delta x = \sqrt{\dfrac{1+4\alpha^4\tau^2}{2\alpha^2}}$,  $\Delta x\rightarrow \sqrt{2}\alpha\tau$ as $\tau\rightarrow\infty$.  The smaller the value of $\alpha$, the smaller the spread as $\tau$  increases.  Reference~\onlinecite{goldberg67} provides a careful discussion of the appropriate choice of parameters given the finite computational space.

\begin{figure}[h]
\centering  
  \resizebox{3.5in}{!}{\includegraphics[angle=-90]{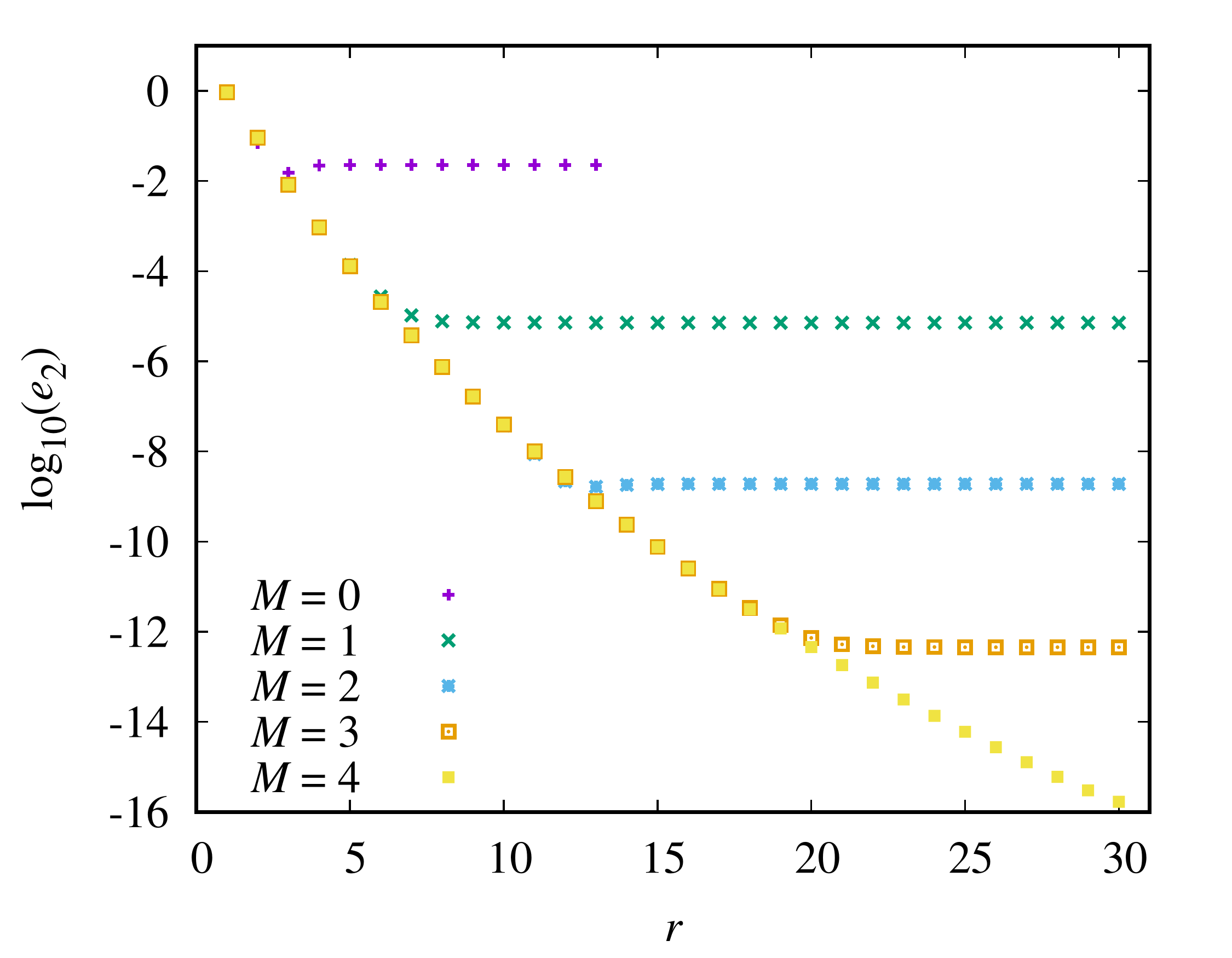}}
\caption{ The error $\log_{10}(e_2)$ as functions of $r$ and $M$ for the free wave packet.  The parameters are given in the text, and $e_2$ is defined in Eq.~(\ref{418}).}
\label{e2_vs_r}
\end{figure}

A sample computation with  $\hbar=m=1$, $\alpha=1$, and $k=2$ using   $x\in[-200,400]$ and $t\in[0,20]$ determines the logarithm of the error as functions of $M$ and $r$ defined by 
\beq
\label{418}
(e_2)^2 = \int_{x_0}^{x_J} dx \ \big|\psi(x,t_\mr{f}) -\psi_\mr{exact}(x,t_\mr{f})\big|^2,
\eeq
where $t_\mr{f}$ is the final time.  These error functions  are displayed at $t=20$ for $\Delta t =0.01$    in Fig.~\ref{e2_vs_r}.

\section{Properties of the algorithm}
\label{sec:04}

\subsection{Stability}
The stability of the algorithm is a measure of the  growth of errors in the numerical solution. To investigate this aspect, we employ von Neumann's stability analysis.  At the $n$th time iteration the local Fourier mode of the solution can be written as 
\beq
\label{401}
\psi_j^n = \xi^n(k,\Delta t, \Delta x)e^{\ts ikx_j}. 
\eeq
Note that the superscript on the growth factor $\xi$ indicates a raising to the $n$th power.  If $|\xi|\leq 1$, the solution does not grow and the algorithm is    stable.  To apply the Fourier mode to Eq.~\eqref{308}, we need
\beq
\dfrac{H\Delta t}{\hbar}\xi^n e^{\ds ikj\Delta x} .
\eeq
We  apply Eq.~(\ref{3.04}) to obtain
\begin{subequations}
\label{402}
\begin{align}
\dfrac{H\Delta t}{\hbar}\xi^n e^{\ds ikj\Delta x} 
&= \dfrac{\Delta t}{\hbar}\left[-\dfrac{\hbar^2}{2m(\Delta x)^2}\ds\sum_{\ell=-r}^r c^{(r)}_\ell e^{ik\ell\Delta x}+V_j\right]\xi^n e^{\ds ikj\Delta x} \\
&= \dfrac{\Delta t}{\hbar}\left[-\dfrac{\hbar^2}{2m(\Delta x)^2}\ds\sum_{\ell=1}^r c^{(r)}_\ell (e^{ik\ell\Delta x}+ e^{-ik\ell\Delta x} - 2)+ V_j\right]\xi^n e^{\ds ikj\Delta x} \\
&=\dfrac{\Delta t}{\hbar}\left[-\dfrac{\hbar^2}{2m(\Delta x)^2}\ds\sum_{\ell=1}^r c^{(r)}_\ell (2\cos{(k\ell\Delta x)} - 2)+ V_j\right]\xi^n e^{\ds ikj\Delta x} \\
&= \dfrac{\Delta t}{\hbar}\left[\dfrac{\hbar^2}{2m(\Delta x)^2}\ds\sum_{\ell=1}^r c^{(r)}_\ell \sin^2\left(\dfrac{k\ell\Delta x}{2}\right)+V_j\right]\xi^n e^{\ds ikj\Delta x} \\
&\equiv b\xi^n e^{\ds ikj\Delta x},
\end{align}
\end{subequations}
where we assume that $V_j$ is small compared to the kinetic energy term, or is constant, so that $b$ does not depend on $j$.   Thus,
\beq
\label{403}
b \approx \dfrac{\hbar\Delta t}{2m(\Delta x)^2}\ds\sum_{\ell=1}^r c^{(r)}_\ell \sin^2\left(\dfrac{k\ell\Delta x}{2}\right). 
\eeq
Furthermore
\begin{subequations}
\label{404}
\begin{align}
B^{(2M,r)}\psi_j^n & = \dfrac{2iH\Delta t}{\hbar} \ds\prod_{s=1}^{2M}\left(1-\dfrac{H\Delta t/\hbar}{z_s^{(2M)}}\right)\xi^ne^{\ts ikj\Delta x} \\
& = 2ib\ds\prod_{s=1}^{2M}\left(1 - \dfrac{b}{z_s^{(2M)}}\right) \xi^ne^{\ts ikj\Delta x}  \\
& = 2ia\xi^n e^{\ts ikj\Delta x},
\end{align}
\end{subequations}
where 
\beq
\label{405}
a =b\prod_{s=1}^{2M} \left(1-\dfrac{b}{z_s^{(2M)}}\right).
\eeq
The quantity $a$ involves the products of complex conjugates and therefore is real.  

By substituting Eqs.~\eqref{401} and \eqref{405} into Eq.~\eqref{308}, we obtain an equation for $\xi$,
\beq
\label{406}
\xi^2+2ia\xi -1 =0,
\eeq
with solutions $\xi=-ia\pm\sqrt{1-a^2}$.  For $a^2<1$, $|\xi|^2 =1$, but for $a^2>1$, there is at least one solution for which $|\xi|^2>1$.\cite{rubin79}  In actual calculations we choose parameters such as $\Delta x$ and $\Delta t$, but also $r$ and $M$, so that $a^2<1$  for stability. 

We use Eq.~\eqref{403} (with    $\hbar=m=1$)   
and evaluate $a$ in  Eq.~\eqref{405} for a range of $k\Delta x\in(0,2\pi)$  and determine the maximum of the ratio $\Delta t/(\Delta x)^2$ for which $a<1$ for all $k$.  The results to two decimals are given in Table~\ref{table:2a}.

\begin{table}[h] 
\renewcommand{\arraystretch}{1.5} 
\begin{tabular}{cccccc}
\hline\hline
$r\backslash M$ & ~~~~0~~~~ & ~~~~1~~~~ & ~~~~5~~~~ & ~~~~10~~~~ & ~~~~15~~~~ \\ 
\hline
 1 & 0.50 & 1.42 & 2.21 & 3.85 & 5.46 \\ 
 2 & 0.37 & 1.06 & 1.66 & 2.89 & 5.04 \\
 3 & 0.33 & 0.93 & 1.46 & 2.55 & 4.44 \\
 4 & 0.30 & 0.87 & 1.36 & 2.37 & 4.07 \\
 5 & 0.29 & 0.83 & 1.29 & 2.26 & 3.20 \\
10 & 0.26 & 0.74 & 1.15 & 2.01 & 3.02 \\
20 & 0.24 & 0.68 & 1.06 & 1.85 & 3.05 \\
30 & 0.23 & 0.65 & 1.03 & 1.79 & 3.13 \\
\hline
\end{tabular}
\caption{Largest value of the ratio $\Delta t/(\Delta x)^2$ for which $a^2<1$ for any value of $k$.}
\label{table:2a}
\end{table}
	    
Table~\ref{table:2a} gives an approximate guide to choosing the step size. For example, if $\Delta x$ is given, we can estimate an appropriate $\Delta t$.  We expect that smaller values than those listed for particular $r$ and $M$ will lead to stable solutions.  The ratio  usually behaves monotonically, but not linearly, with increasing $M$ and $r$. The values in Table~\ref{table:2a} are a rough guide because the effect of the potential is not included.

\subsection{Unitarity} 

As with many explicit methods, normalization is not rigorously preserved.\citep{askar78,leforestier91,kosloff83}  In this section we explore the deviation from the exact norm.
  In fact, we may consider more generally the error of the moments or expectation values of the Hamiltonian raised to some power, i.e., $H^s, s=0,1,\dots, M$ for some value of $M$.  Consider 
\beq
\label{407}
\sin\left(H\Delta t/\hbar\right) = S_{2M}\left(H\Delta t/\hbar\right) + {\cal   O} \left[\left(H\Delta t/\hbar\right)^{2M+3}\right],
\eeq
where 
\begin{subequations}
\label{2.04}
\begin{align}
S_{2M}(z) & \equiv z\left[1-\dfrac{1}{3!}z^2 + \dfrac{1}{5!}z^4
 - \cdots + \dfrac{(-1)^M}{(2M+1)!}z^{2M}\right] \\
 & = z\ds\prod^{2M}_{s=1}\left(1-\dfrac{z}{z^{(2M)}_s}\right).
\end{align} 	
\end{subequations}
 We express our algorithm as 
\beq
\label{408}
\psi^{n+1} = \psi^{n-1} -2iS_{2M}\left(H\Delta t/\hbar\right)\psi^n.
\eeq
Because $\ds\psi^{n-1}=e^{\ts iH\Delta t/\hbar}\psi^n $ exactly, 
\beq
\label{409}
\psi^{n+1} = \left[e^{\ts iH\Delta t/\hbar} - 2iS_{2M}(H\Delta t/\hbar)\right]\psi^n.
\eeq
Hence
\beq
\label{410}
\begin{array}{ll}
\left\langle \psi^{n+1} \big|H^s\big|\psi^{n+1}\right\rangle = & \left\langle\psi^n\bigg|\left[e^{\ts -iH\Delta t/\hbar} +2iS_{2M}\right]H^s\right.  \\  
& ~~~~~~\left. \times\left[e^{\ts iH\Delta t/\hbar} -2iS_{2M}\right]\bigg|\psi^n\right\rangle,
\end{array}
\eeq
 and   it follows that
\beq
\label{411}
\left\langle \psi^{n+1} \big|H^s\big| \psi^{n+1}\right\rangle 
= \left\langle\psi^n\big|H^s\big|\psi^n\right\rangle + {\cal O}\left[(\Delta t)^{2M+4}\right],
\eeq
where we have used the expression for $S_{2M}$ from Eq.~\eqref{407}.  Thus the normalization and energy (and higher moments) are constant to  order  $(\Delta t)^{2M+4}$, a result that is consistent with Ref.~\onlinecite{askar78} for $M=0$. 

\subsection{Error and computation time}
The efficacy of the algorithm can be expressed in terms of the computation time required to obtain a solution with the specified maximum error.  
The numerical solutions are  obtained by truncating the Taylor series in $x$ and $t$.  By assuming that the series in one variable has a much smaller error than the other variable, we consider the respective truncation errors separately.  When necessary, we can interpolate to situations where both types of errors contribute similar amounts.

At a given time $t$ the truncation error of the spatial integration with the $r$th-order expansion is
\beq
\label{412}
e^{(r)} \approx C^{(r)}(\Delta x)^{2r},
\eeq
where we assume $C^{(r)}$   varies slowly with $r$.  The spatial step size can be adjusted to obtain an acceptable error by making an appropriate choice of $J$, because
\beq
\label{413}
\Delta x = \dfrac{x_J-x_0}{J} \approx \left(\dfrac{e^{(r)}}{C^{(r)}}\right)^{1/2r} \mbox{ so that }  e^{(r)} \approx \dfrac{\mr{const}}{J^{2r}}.
\eeq
The CPU time is proportional to the number of basic computer operations.  The largest number of operations are due to the iterated time advances. The matrix $B^{(2M,r)}$ is calculated at the beginning once and for all, and is a symmetric banded $J\times J$ matrix with $(4M+3)r$ bandwidth.  For a particular time step $\Delta t$ and value of $M$ 
\beq
\label{414}
\mbox{CPU time} \propto Jr \propto \dfrac{r}{(e^{(r)})^{1/2r}}.  
\eeq
By minimizing Eq.~(\ref{414}) with respect to $r$, the minimum CPU time occurs when 
 \beq
 \label{415}
 r\approx -\dfrac{\ln e^{(r)}}{2}.
 \eeq
 Similarly for a fixed $r$ and $\Delta x$, the truncation  error due to a nonzero value of $\Delta t$ is
 \beq
 \label{416}
 e^{(2M)} \approx C^{(2M)}(\Delta t)^{2M+3},
\eeq
where again $C^{(2M)}$ is assumed to be slowly varying with $M$.  Because the expansion of $\sin(H\dt/\hbar)$ has factors involving $\dt/z^{(2M)}_s$ rather than $\dt$, we replace $\dt$ by $\dt/z_{\rm{avg}}^{(2M)}\approx \dt/M$.  If we take the final time to be fixed at $N\dt$, then
\beq
\label{417}
\mbox{CPU time} \propto \frac{1}{M}\left(e^{(2M)}\right)^{-1/(2M+3)}.
\eeq
The CPU time is a monotonically decreasing function of increasing $M$. 

The error when an exact solution is available  is expressed as a Euclidean norm using Eq.~(\ref{418}). 
For the case when no exact solutions are available, a possible definition is 
\beq
\label{419}
\left(\eta_2^{(2M,r)}\right)^2 = \int_{x_0}^{x_J} dx \ \big|\psi^{(2M+2,r+1)}(x,t_\mr{f}) - \psi^{(2M,r)}(x,t_\mr{f})\big|^2
\eeq
as an estimate of the error.

\section{Numerical examples}\label{sec:05}

\subsection{Oscillating and pulsating harmonic oscillator wave functions}

Consider the potential function for the one-dimensional
harmonic oscillator,
\beq
\label{501}
V (x) = \dfrac{1}{2}m \omega^2 x^2.
\eeq
A wave function that satisfies the \SE~is~\cite{vandijk14}
\beq
\label{502}
\renewcommand{\arraystretch}{2}
\begin{array}{ll}
\psi_n (\alpha,\beta; x,t) = &
\left(\dfrac{\alpha^2\beta}{\sqrt{\pi}2^n n!}\right)^{1/2} \dfrac{e^{\ts -i(n+1/2)\theta}}{f^{1/4}} \\
& \times H_n(\xi)e^{\ts -\xi^2/2+i{\cal T}},
\end{array}
\eeq
where
\begin{subequations}
\label{503}
\begin{align}
\alpha& =\sqrt{m\omega/\hbar},  \quad  f = \alpha^4\cos^2\omega t + \beta^4\sin^2\omega t  \\
\xi & =  \beta[\alpha^2(x-A\cos\omega t)-k\sin\omega t]/f^{1/2}\\
{\cal T} & = \dfrac{1}{2f}\Big\{\alpha^2[(\beta^4-\alpha^4)x^2-k^2+\beta^4A^2]\sin\omega t\cos\omega t  \nonumber \\
&{} + 2[\alpha^4 kx\cos\omega t + \beta^4A(k\sin\omega t-\alpha^2 x)\sin\omega t]\Big\}  \\
\theta & =  \arctan\left(\dfrac{\beta^2\sin\omega t}{\alpha^2\cos\omega t}\right) + 2\pi\nu, \quad  \nu = \left\lfloor\dfrac{\omega t + \pi}{2\pi}-\epsilon\right\rfloor,
\end{align} 
\end{subequations}
and $H_n(\xi)$ is the $n$th order Hermite polynomial.\footnote{In the definition of $\theta$ the $\arctan(Y/X)$ function gives the polar angle coordinate between $-\pi$ and $\pi$ of the Cartesian point $(X,Y)$.  In computer languages this function is often denoted as {\tt atan2(X,Y)}.  The infinitesimal positive $\epsilon$ in the definition of $\nu$ ensures that the phase $\theta$ is a continuous function of $t$.} 
This wave function gives rise to oscillating and pulsating wave packets that oscillate with frequency $\hbar\alpha^2/(2\pi m)$ and pulsate with frequency $\hbar\beta^2/(2\pi m)$.   Because the $n$th order Hermite polynomial has $n$ zeros,  the wave function and the corresponding packet also have $n$ zeros. The width of the wave function  compresses and expands with the pulsating frequency.  The initial ($t = 0$) wave
function is the $n$th-energy eigenstate of the particle subject to
an oscillator characterized by $\beta$, rather than $\alpha$, displaced from the origin by amount $A$ and with  momentum $\hbar k$.

\begin{figure}[htpb]
  \centering
   \resizebox{3.5in}{!}{\includegraphics[angle=-90]{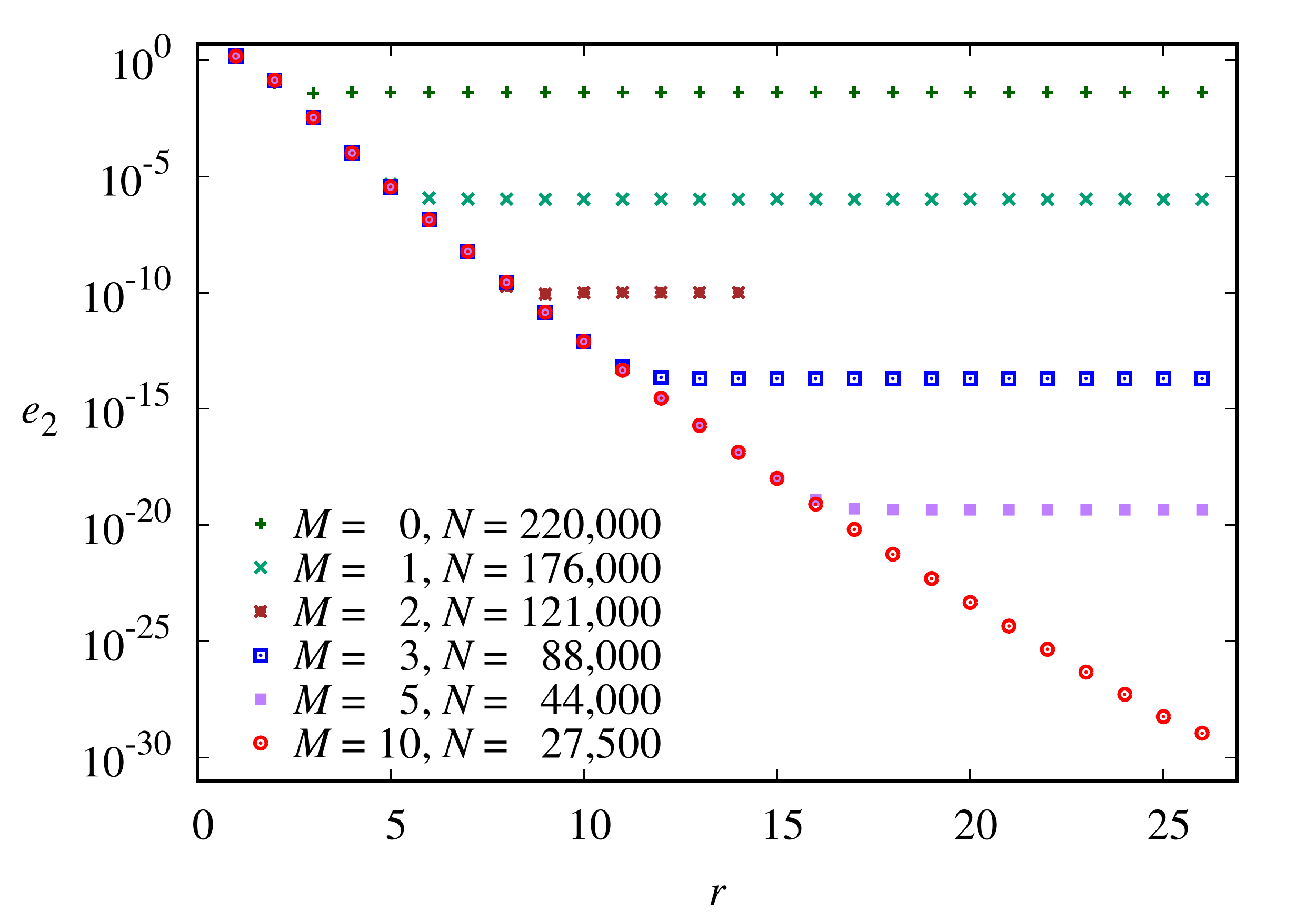}}
\caption{The error defined in Eq.~\eqref{418} as a function of $r$ for various values of $M$.    Time ranges from 0 to $110\pi$ and the number of time steps $N$ is indicated for each value of $M$. 
$\Delta x = 0.1$ in all cases.}
\label{fig:02a}
\end{figure}

To obtain an overview of the scope of the method, we plot the errors of the numerical solutions  as functions of $r$ and $M$ in Fig.~\ref{fig:02a} with     
$n=4$, $\alpha = \sqrt{0.2}$, $\beta = 2\alpha$, $k=1$, $A=10$,  $J=1600$, and  $t\in [0,110\pi]$ and $x\in [-80,80]$, respectively.  In this example we set  $\hbar=m=1$. 

The calculations are done with methods I and II  and give identical results.  For $M=2$   the results are not shown for $r>16$ because the algorithm is unstable, which illustrates the restrictions on the stability criteria when $r$ increases as shown in Table~\ref{table:2a}. 

For this  example the error  compared to the exact solution can be calculated.   To test the reliability of $\eta^{(2M,r)}$ as an estimate of the error, we compared $e_2$ and $\eta^{(2M,r)}$, and observed that the two quantities are nearly equal, the latter yielding a better than order-of-magnitude  estimate. The instability of the algorithm is indicated by the deviation of the normalization  from unity. 
   
To investigate the efficacy of the method, we determine the CPU time to compute the wave function with a modest error $e_2\lessapprox 10^{-3}$ for different values of $r$ and $J$ when $M=10$ and $\Delta t=\pi/120$.  The results are given in Table~\ref{table_2}, which shows that  the most efficient calculation occurs at $r\approx 7$.

\begin{table}[h]
\renewcommand{\baselinestretch}{1.5}
\begin{tabular}{cccccc}
\hline\hline 
~~$M$~~ & ~~~$r$~~~ & ~~~$J$~~~ & ~~~~$\Delta t$~~~~ & ~~~$e_2$~~~ & ~~~~CPU(s)~~~~ \\ 
\hline
10 & 3 & 750 & $\pi/120$ & $1.79\times 10^{-3}$ & 13.08 \\  
• & 4 & 506 & • & $1.00\times 10^{-3}$ & 10.35 \\  
• & 5 & 381 & • & $1.00\times 10^{-3}$ & 9.02 \\  
• & 6 & 318 & • & $9.71\times 10^{-4}$ & 8.56 \\  
• & 7 & 280 & • & $9.64\times 10^{-4}$ & 8.46 \\  
• & 8 & 255 & • & $9.58\times 10^{-4}$ & 8.50 \\  
• & 9 & 237 & • & $9.76\times 10^{-4}$ & 8.68 \\  
• & 10 & 224 & • & $9.71\times 10^{-4}$ & 8.92 \\  
• & 15 & 190 & • & $9.30\times 10^{-4}$ & 10.574 \\  
• & 20 & 175 & • & $9.36\times 10^{-4}$ & 12.60 \\ \hline 
\end{tabular}
\caption{The CPU time as a function of $r$ and $J$ giving an error
$e_2\lessapprox 10^{-3}$. The CPU time depends on the computer used and  is given for comparative purposes only.}
\label{table_2}
\end{table}

\begin{figure}[htpb]
  \centering
   \resizebox{3.5in}{!}{\includegraphics[angle=-90]{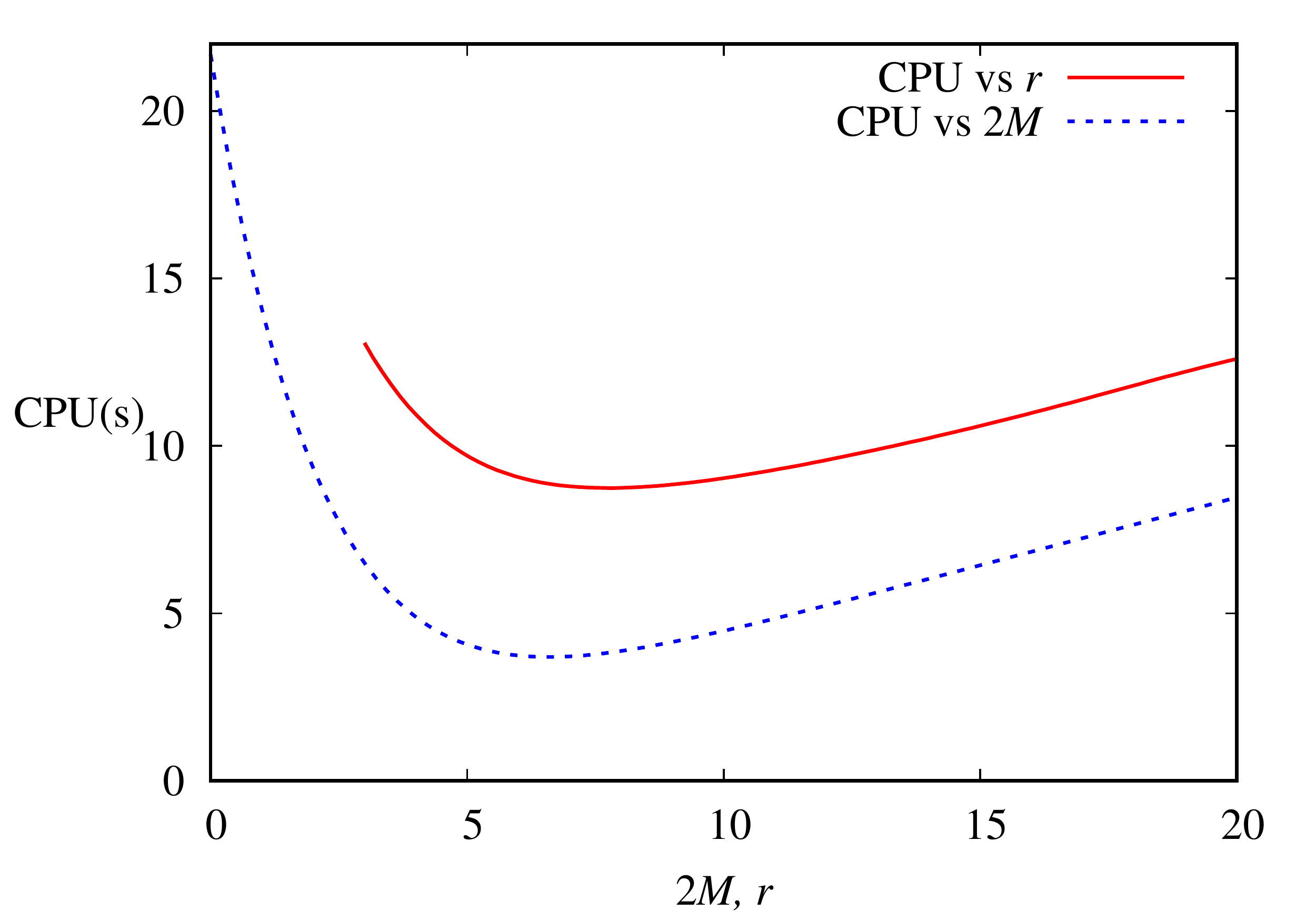}}
\caption{The CPU time as a function of either $M$ or  $r$ with the other variable held constant for the error  $e_2 = 10^{-3}$.  Note the minimum CPU time as a function $r$ which is expected.  The minimum CPU time as a function $M$ is due to the fact that for $M>3$, the error is due to spatial integration and larger $M$ merely results in a larger number of basic operations. }
\label{fig_ex_1_2}
\end{figure}

We repeat this process varying $M$ and $\Delta t$ (or $N$) and list the results in Table~\ref{table_3}.
\begin{table}[h]
\begin{tabular}{cccccc}
\hline\hline
~~$M$~~ & ~~~$r$~~~ & ~~~$J$~~~ & ~~~$\Delta t$~~~ & ~~~~~$e_2$~~~~~ & ~~~CPU(s)~~~ \\  
\hline 
0 & 7 & 280 & ~~~$\pi/7280$~~~ & 9.97$\times 10^{-4}$ & 21.69 \\  
1 &   &     & $\pi/280$ & 9.86$\times 10^{-4}$ & 2.73 \\ 
2 &   &     & $\pi/160$ & 8.56$\times 10^{-4}$ & 2.64 \\  
3 &   &     & $\pi/120$ & 9.64$\times 10^{-4}$ & 2.79 \\  
4 &   &     & $\pi/120$ & 9.64$\times 10^{-4}$ & 3.60 \\  
5 &   &     & $\pi/120$ & 9.64$\times 10^{-4}$ & 4.41 \\ 
7 &   &     & $\pi/120$ & 9.64$\times 10^{-4}$ & 6.03 \\ 
10 &  &     & $\pi/120$ & 9.64$\times 10^{-4}$ & 8.46 \\ \hline
\end{tabular}
\caption{The CPU time as a function of $M$ and $\Delta t$ (or $N$) yielding an error $e_2\lessapprox 10^{-3}$.  The values of $r$ and $J$ are the ones giving the smallest CPU time in Table~\ref{table_2}.} 
\label{table_3}
\end{table}
The behavior of the CPU time as functions of $r$ and $M$ is displayed in Fig.~\ref{fig_ex_1_2}.  As expected (see Sec.~IV.C), the CPU time versus $r$ has a minimum at $r\approx 7$.  We expect the CPU time to be a decreasing function of $M$, but from Table~\ref{table_3} we see that for $M \ge 3$,  the error is constant.  It seems  that the temporal error decreases until it matches the spatial error which then becomes dominant and unchanging for larger $M$.  The linear increase of the CPU time versus $M$ for $M>3$  is due to having a constant error and additional computations as $M$ is made larger.

The behavior of the CPU time as a function of $M$ is illustrated more incisively by examining the behavior for a small error, e.g., $e_2\approx 10^{-10}$ (see Table~\ref{table_5}).   There is a significant increase in efficiency for small values of $M$, but for larger $M$, the efficiency is approximately constant.

\begin{table}[h]
\begin{tabular}{cccccc}
\hline\hline
~~$M$~~ & ~~~$r$~~~ & ~~~$J$~~~ & ~~~$\Delta t$~~~ & ~~~~~$e_2$~~~~~ & ~~~CPU(s)~~~ \\  
\hline 
0 & 20 & 1600 & ~~~$\pi/2,000,000$~~~ & 1.21$\times 10^{-10}$ & 2547 \\  
1 &   &     & $\pi/3,800$ & 9.89$\times 10^{-11}$ & 5.54 \\ 
2 &   &     & $\pi/880$ & 4.80$\times 10^{-11}$ & 1.57 \\  
3 &   &     & $\pi/364$ & 1.93$\times 10^{-10}$ & 0.88 \\  
5 &   &     & $\pi/298$ & 7.33$\times 10^{-10}$ & 1.07 \\ 
10 &  &     & $\pi/120$ & 1.27$\times 10^{-11}$ & 0.94 \\ \hline
\end{tabular}
\caption{The CPU time as a function of $M$  for   $e_2\approx 10^{-10}$. There is a significant gain in efficiency (reduction of CPU time) as $M$ varies from 0 to 3, but there is no significant additional gain for larger $M$.}
\label{table_5}
\end{table}

The error estimate $\eta_2^{(2M,r)}$ is found  to be a good approximation of $e_2$; in most cases the difference is less than 10\%. An important and helpful feature for determining the stability is the behavior of the  normalization.   Whenever it is close to the original value, the procedure is stable, although that does not guarantee a small error.

\subsection{Decay through a barrier}
\label{barrier}
The    decay of a particle through a strong barrier can be simulated using the potential function 
\beq
\label{504}
V(x) = \begin{cases}
\infty~   & (x<0) \\
\dfrac{\lambda}{w\sqrt{\pi}}e^{\ts -(x-a)^2/w^2} & (x\ge 0),
\end{cases}
\eeq
with $a$ and $w >0$.
As the parameter $w$ is shrunk to zero, the potential reduces to the $\delta$-function potential $V(x) = \lambda\delta(x-a)$.  This limiting case has been used to obtain exact wave functions for nuclear decay problems.\cite{vandijk02}  We choose the initial state to be the $n$th normalized eigenstate of the infinite square well of width $a$, i.e.,
\beq
\label{505}
\psi_n(x)= \begin{cases}
\sqrt{\dfrac{2}{a}}\sin\left(\dfrac{n\pi x}{a}\right) &  (0\le x \le a) \\
0 & \mbox{otherwise.}
\end{cases}
\eeq
Imagine a particle bound in the infinite square well up to time $t=0$, after which the infinite potential step at $x=a$ is replaced by Eq.~\eqref{504} and the particle is allowed to escape.  The initial state approximates a resonant state of the system.

\begin{figure}[h]
\centering
   \resizebox{3.5in}{!}{\includegraphics[angle=-90]{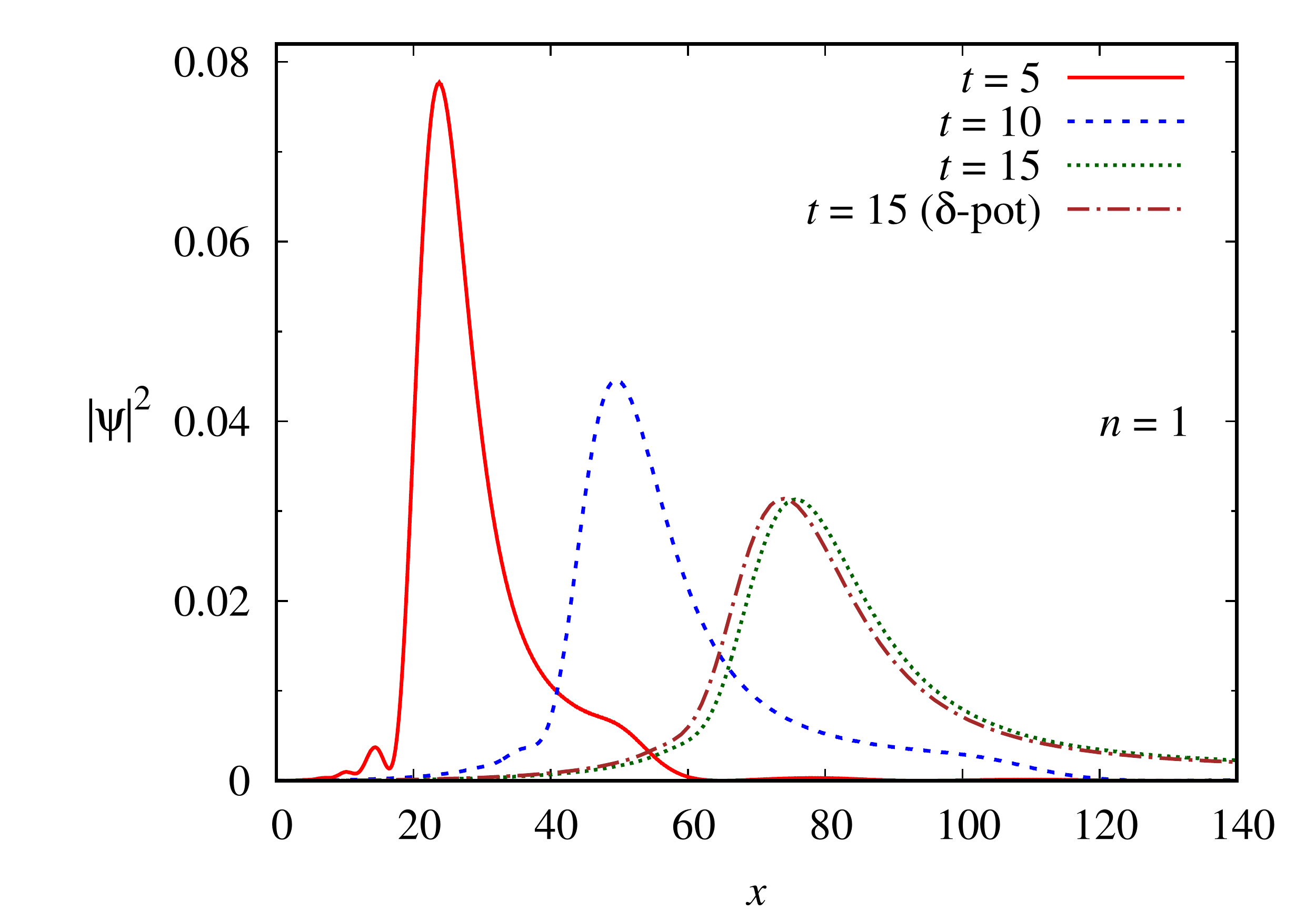}}
\caption{The magnitude-squared of the wave function at times $t=5, 10$, and 15 for $\lambda=3$, $w=1/10$, $a = 1$, and $n=1$ with $\Delta x =2/15$, $\Delta t=1/70$ and $2M=28,r=29$.  For $t=15$ the exact wave function when $w\rightarrow 0$ is also shown.
}
\label{fig_ex_2a_n1}
\end{figure}

 To study a specific case we adopt  $\hbar=2m=1$ and the parameters, $\lambda =3$, $w = 0.1$, $a = 1$, and $n=1$, and compute the wave function at times $t=0$, 10, and 15.  The numerical calculations are done over the computational space $x\in[0,800]$ with $J=6000$  so that $\Delta x= 2/15$ and $\Delta t=1/70$.  The choice of $2M=28,r=29$ leads to stable and efficient solutions.    
The magnitude of the wave function is displayed in Fig.~\ref{fig_ex_2a_n1}.  The estimated error at $t=15$ is $\eta_2^{(28,29)} = 0.0054$.

\begin{figure}[h]
\centering
   \resizebox{3.5in}{!}{\includegraphics[angle=-90]{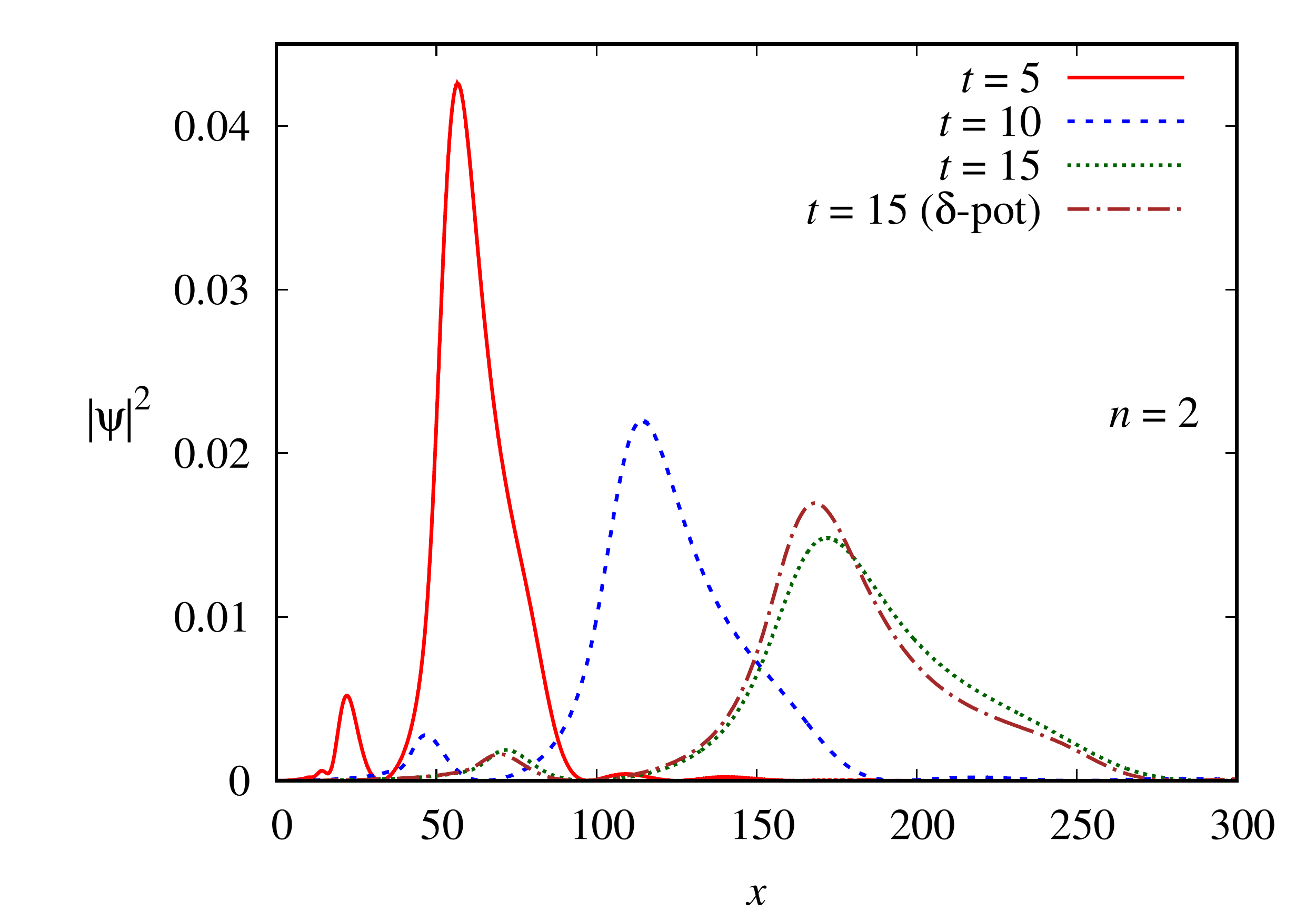}}
\caption{The magnitude-squared of the wave function at times $t=5, 10$, and 15 with the parameters, except for $n=2$,  the same as in Fig.~\ref{fig_ex_2a_n1}.   The estimated error is $\eta_2^{(28,29)}= 0.0285$.
}
\label{fig_ex_2a_n2}
\end{figure}

When the initial state is characterized by $n=2$, the main peak of the wave packet travels faster as is shown in Fig.~\ref{fig_ex_2a_n2}; note the larger range of $x$ values.

As the $n=2$ state decays it produces a wave function outside the barrier which travels with a higher speed commensurate with the energy of the resonant state, but  also makes a transition to the $n=1$ ground state  state, which then decays.  Hence we see smaller peaks at distances similar to the position of the peaks in Fig.~\ref{fig_ex_2a_n1}. We explore this behavior further by considering the escape probability $P(t)$ and the survival probability $S_n(t)$ of the $n$th initial state, which are defined, respectively, as
\begin{subequations}
\label{506}
\begin{align}
P(t) & = \ds\int_0^a \; dx |\psi(x,t)|^2 \\
S_n(t) & = \dfrac{2}{a}\ds \left| \int_0^a \; dx \sin\left(\dfrac{n\pi x}{a}\right)\psi(x,t) \right|^2.
\end{align}
\end{subequations}

\begin{figure}[h]
\centering
   \resizebox{3.5in}{!}{\includegraphics[angle=-90]{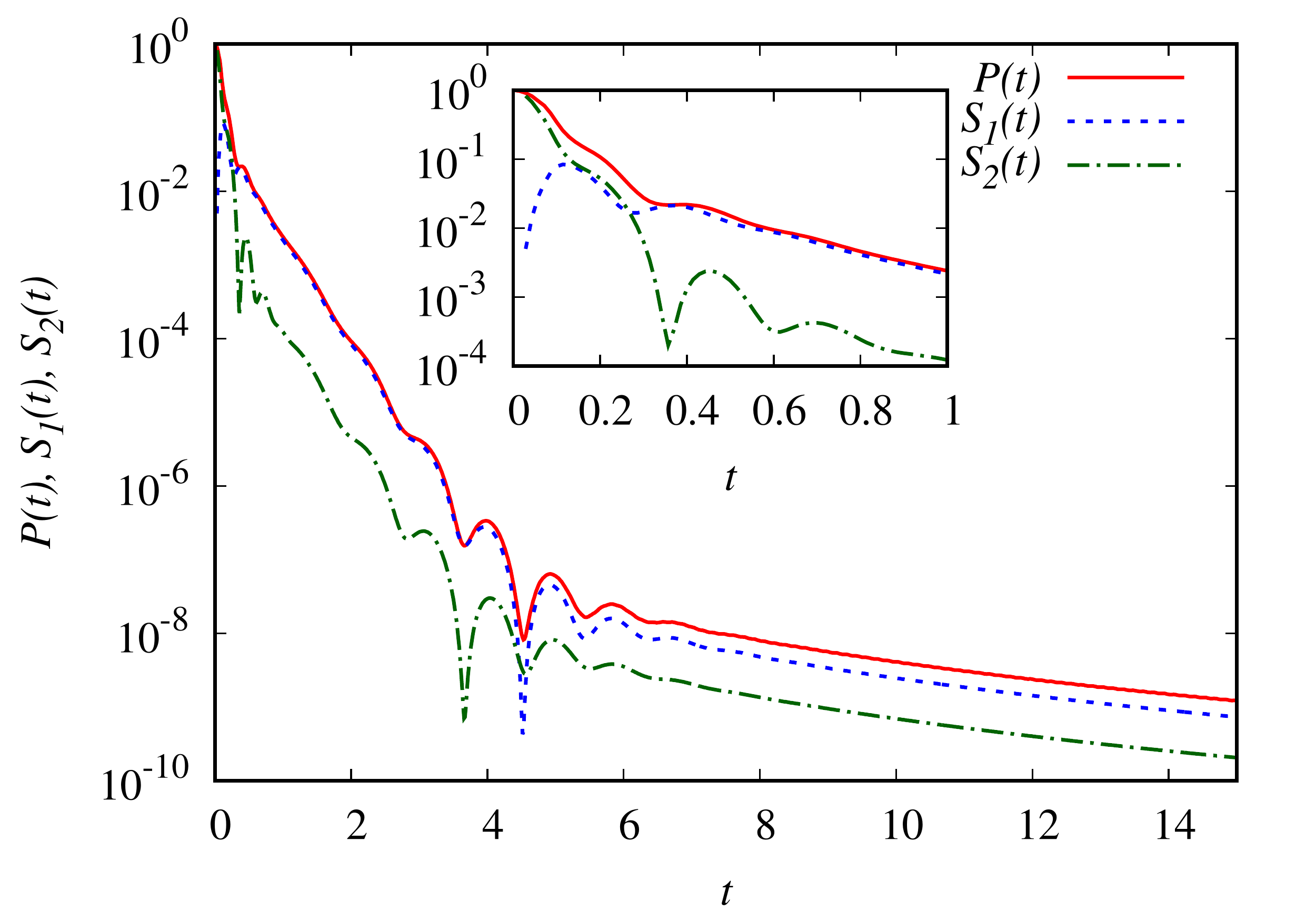}}
\caption{The escape and survival probabilities  as function of time for the parameters, of Fig.~\ref{fig_ex_2a_n1} except for $n=2$.
}
\label{fig_ex_2b}
\end{figure}

The nonescape probability is the probability that the decaying particle is in the region $0~<x\le~a$ at time $t$, and the survival probability $S_n(t)$ is the overlap of the $n$th original square-well state component with the wave function at time $t$.  We plot these probabilities for  $n= 2$  in Fig.~\ref{fig_ex_2b} and make the following observations.

\begin{enumerate}
\item There are two exponential decay regions: one for very short times when the $n=2$ state decays and another when the $n=1$ state decays at a slower rate.

\item As the $n=2$ state decays, its survival probability decreases, increasing the probability of finding the particle outside the potential region and initially increasing $S_1(t)$.

\item The $n=1$ state decays slower than the $n=2$ state, so that after the $n=2$ is depleted the $n=1$ continues to decay at the reduced rate.

\item The survival probabilities are smaller than the nonescape probabilities because there are (higher-energy) components that could be part of the initial wave function.  The nonescape and survival probabilities  track each other.
\item The long time behavior of all three probabilities in Fig.~\ref{fig_ex_2b} goes as $t^{-3}$.
\item There is a hint in the plot that the very short-time behavior of the probabilities is proportional to $t^2$.
\item During the transitional time periods when the probability profiles change from quadratic to the first exponential, and then to the second exponential, and to the $t^{-3}$ decay, interference effects cause fluctuations  of the probabilities in time.\cite{winter61}  During these fluctuations there are times that the probability of the decaying particle being inside the barrier actually increases.  This counterintuitive phenomenon is referred to as quantum backflow and was studied  recently.\cite{bracken94}  See also Refs.~\onlinecite{nogami00,vandijk19}.

\end{enumerate} 

The accuracy of the calculations of this section is limited because of the constraint of the wave $\psi(0,t)=0$.  Such boundary conditions can be included using the summation-by-parts-simultaneous approximation term  technique.\cite{mattsson04,nissen13}  However, there is no general formulation which applies for an arbitrary value of $r$, and considering particular cases is beyond the scope of this paper. 

\subsection{Hermite-Gaussian packets}
\subsubsection{Free packets in one dimension}
Free pulsating packets are described in Refs.~\onlinecite{vandijk14,vandijk20} and references therein.  In one dimension the wave function is 
\beq
\label{5c.01}
\renewcommand{\arraystretch}{3}
\begin{array}{ll}
\psi_n(\alpha;x,t)   = \dfrac{N_n(\alpha)e^{\ts in\theta}}{\sqrt{1+2i\alpha^2 \tau}}H_n(\xi)\exp(-\xi^2/2) \\
\times \exp\left\{i\left[\dfrac{2\alpha^4\tau(x-A)^2 + 2k_0(x-A) - 2k_0^2\tau}{2(1+4\alpha^4\tau^2)}\right]\right\},
\end{array}
\eeq
where
\begin{subequations}
\label{5c.02}
\begin{align}
\tau & = \hbar(t-t_0)/(2m),  \qquad \theta = -\arctan(2\alpha^2\tau)&\\
\xi & = \dfrac{\alpha[(x-A)-2k_0\tau]}{\sqrt{1+4\alpha^4\tau^2}}, \quad
 N_n(\alpha) = \sqrt{\dfrac{\alpha}{\sqrt{\pi}2^nn!}}.&
\end{align}
\end{subequations}
The integer $n=0,1,\dots$ labels a wave function that, at $\tau=0$ or $t=t_0$, is the $n$th energy state of the harmonic oscillator centered at $x=A$ with oscillator constant $\hbar^2\alpha^4/m$.

As an overview of the efficacy determined by the expansion of the time-evolution operator, we survey the accuracy of the numerical calculations over a range of values of $M$ holding $\Delta x$, $J$, $t_0$, and the maximum time constant for a set of parameters of the free wave packet.  The parameters are $\hbar=2m=1$, $\alpha = 0.4$, $A=0$, $k_0=2$, $t_0=20$, $\Delta x =0.15$, $J = 4000$, $r= 10$, and $t_\mr{final} = 50$;  the computational space is defined by $x\in[-300,300]$.  We tabulate the largest $\Delta t$ (smallest ($1/\Delta t$) that yields a stable algorithm and the corresponding CPU time and error in Table~\ref{table_4a}.

\begin{table}[h]
\begin{tabular}{cccccc}
\hline\hline
$M$ & $1/\Delta t$ &$\eta_2^{(2M,10)}$ &$e_2$ & CPU(s) \\
\hline
0 & 5000 & ~~$1.57\times 10^{-2}$~~ & ~~$1.57\times 10^{-2}$~~ & 654\\
1 & 500 & $3.10\times 10^{-7}$ & $3.10\times 10^{-7}$ &127 \\
2 & 250 & $2.08\times 10^{-10}$ & $4.29\times 10^{-10}$ & 94 \\
3 & 89 & $4.17\times 10^{-12}$ & $3.76\times 10^{-11}$ & 45 \\
4 & 78 & $3.83\times 10^{-10}$ & $5.37\times 10^{-10}$ & 49 \\
5 & 77 &  $3.02\times 10^{-12}$ & $3.75\times 10^{-10}$ &57  \\
10 & 45 & $3.51\times 10^{-12}$ & $3.75\times 10^{-10}$ & 61 \\
14 & 32 & $3.11\times 10^{-12}$ & $3.75\times 10^{-10}$ & 59 \\
\hline\hline 	
\end{tabular}
\caption{Errors of the free Hermite-Gaussian packet propagation.  The parameters  are given in the text. Except for the first three lines, $1/\Delta t$ is the smallest integer that yields a stable algorithm.}
\label{table_4a}
\end{table}

\subsubsection{Scattering from a P\"oschl-Teller barrier}
We consider the example discussed in Ref.~\onlinecite{vandijk20} of the Hermite-Gaussian wave function scattering from the P\"oschl-Teller barrier,\cite{flugge74,landau58a} 
\beq
\label{5c.03}
V(x) = \dfrac{V_0}{\cosh^2(\mu x)}.
\eeq
The  transmission and reflections coefficients calculated from the time-independent Schr\"odinger equation are
\beq
\label{5c.04}
T^\mr{(PT)}(k)=\dfrac{q^2}{1+q^2}, \quad R^\mr{(PT)}(k) = \dfrac{1}{1+q^2},
\eeq
where
\beq
\label{5c.05}
q = \dfrac{\sinh(\pi k/\mu)}{\cosh\left({\dfrac{\pi}{2}\sqrt{4V_0/\mu^2 -1}}\right)}.
\eeq
The transmission coefficient is the long-time probability of the transmitted wave packet 
\beq
\label{5c.06}
T = \lim_{t\rightarrow\infty}\int_0^\infty \ dx \ |\psi(x,t)|^2 .
\eeq   
The P\"osch-teller Transmission coefficient  $T^\mr{(PT)}(k)$ depends on  a single value of $k$, whereas $T$ in Eq.~\eqref{5c.06} depends on a (narrow) band of $k$ values around a central value.

We consider an example with   $\hbar=2m=1$, $V_0 = 10$ and $\mu =2$ for  $n=2$, $\alpha = 0.4$, $k = 3$, A$=0$, and $t_0=10$ for the incident wave function; and $x\in[-300,300]$, $\Delta x = 0.075$, $\Delta t = 0.0025$, $t_{\max} = 30$, $r = 10$, and $M=14$.

\begin{figure}[h]
\centering
\resizebox{3.5in}{!}{\includegraphics[angle=-90]{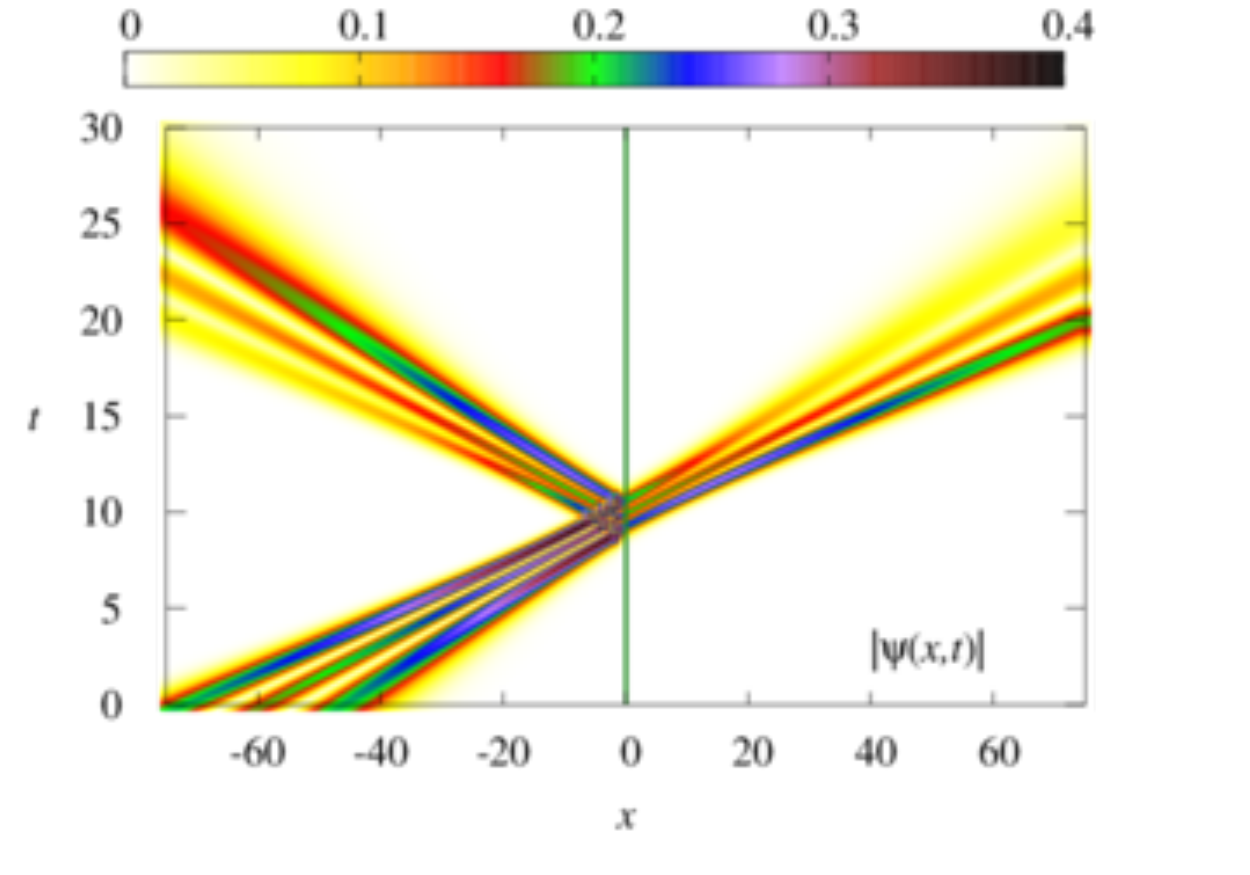}}
\caption{(Color online) The color-coded magnitude of the time-dependent Hermite-Gaussian wave function of particle scattering from the P\"oschl-Teller potential.  The parameters are given in the text.  The center of the potential is indicated by the green vertical line.
}
\label{fig_ex3_a}	
\end{figure}

\begin{figure}[h]
\centering
\hspace{-0.3in}\resizebox{3.7in}{!}{\includegraphics[angle=-90]{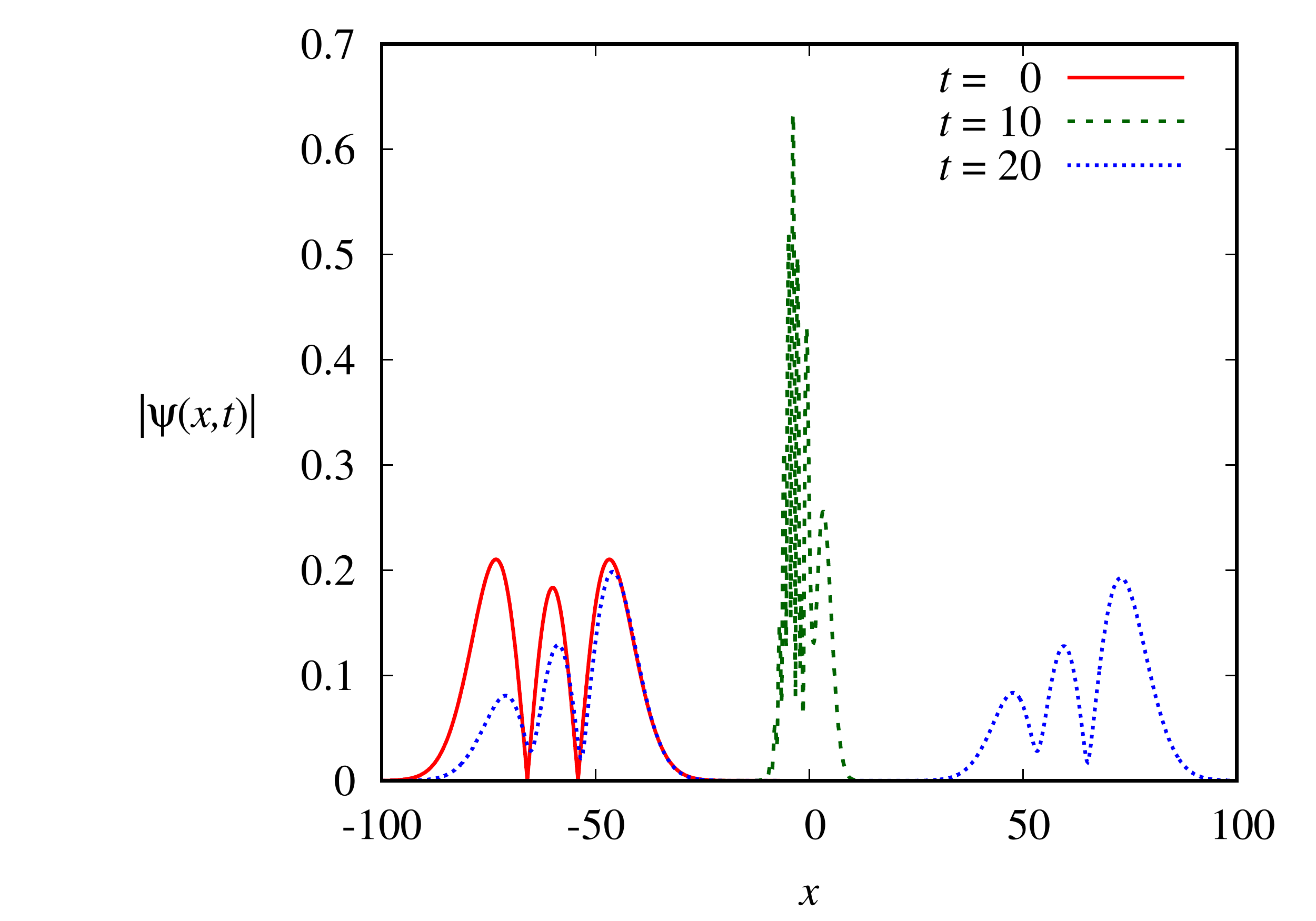}}
\caption{The magnitude of the wave function as a function of position with the parameters of Fig.~\ref{fig_ex3_a}.}
\label{fig_ex3_b}
\end{figure}

Figure~\ref{fig_ex3_a} displays the magnitude of the wave function as a function of $x$ and $t$ as the wave travels to the scattering region and away from it at a later time.  The magnitude of the wave function at three different times is plotted in Fig.~\ref{fig_ex3_b}.

\begin{figure}[h]
\centering
\resizebox{3.5in}{!}{\includegraphics[angle=-90]{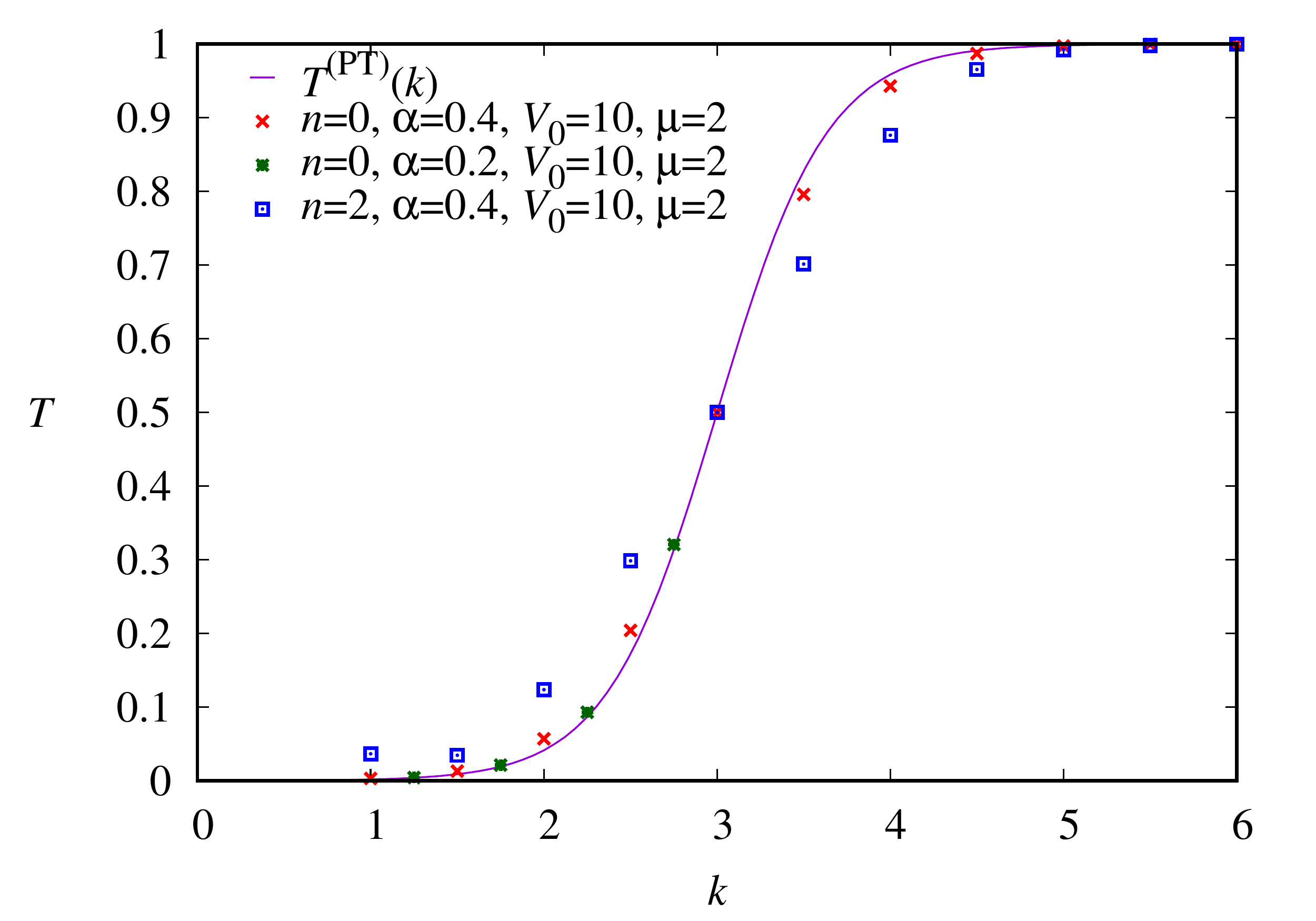}}
\caption{The transmission coefficient as a function of the wave number $k$ for the P\"osch-Teller potential calculated using (a) the time-independent method Eq.~\eqref{5c.04} (solid line) and (b) the asymptotic time-dependent wave functions Eq.~\eqref{5c.06} (points).  In all cases $V_0=10$ and  $\mu = 2$ and $n$ and $\alpha$ of the time-dependent initial wave function are indicated.} 
\label{fig_ex3_c}
\end{figure}

We calculate the transmission coefficient obtained analytically by the time independent method, i.e., Eq.~\eqref{5c.04}, and by using the time-dependent wave functions at a large time, i.e., Eq.~\eqref{5c.06}.  The results are plotted in Fig.~\ref{fig_ex3_c}.

Because the uncertainty of the momentum of the free Hermite-Gaussian wave function is $(\Delta p)_n = \hbar\alpha\sqrt{n+\frac{1}{2}}$ and is constant, we expect greater deviation from the time-independent transmission coefficients as $n$ and $\alpha$ take on larger values.

We enumerate a number of observations. 
\begin{enumerate}

\item The width of the free (unscattered) Hermite-Gaussian wave packet is a  minimum  when $t=t_0$; when $t - t_0$ becomes more negative or more positive the packet spreads.  Thus as $t$ increases when $t-t_0  <0$, the packet becomes narrower,  and when $t-t_0>0$, it spreads as shown in Fig.~\ref{fig_ex3_a}.  

\item The free Hermite-Gaussian wave function has nodes at the zeros of the Hermite polynomials which occur closer to (farther from) each other  as $|t-t_0|$ becomes smaller (larger).  This behavior can be seen in Fig.~\ref{fig_ex3_a} in the lighter bands embedded in the darker ones.

\item It has been suggested that wave packets that are transmitted through a barrier approximately retain the shape of the incident packet.\cite{low98}  This behavior seems a good approximation for square barriers, but features such as extrema are preserved as well in the transmission through a smoothly varying potential such as P\"oschl-Teller. See Fig.~\ref{fig_ex3_b} where the transmitted packet has the same number of minima and maxima as the incident packet although their amplitudes are different.

\item By examining the  amplitudes of the transmitted and reflected wave packets, we note that the transmitted packet is weighted toward the faster components of the packet, whereas the reflected packet is weighted toward the slower components.  This filtering effect of the higher  momentum components was discussed in Refs.~\onlinecite{dumont93,dumont20}.
\end{enumerate}

\section{Corollary implications}\label{sec:06}

\subsection{Hermitian operator}
We now consider variants of the proposed  algorithm that allow us to generalize integration methods that have been found practical.  We have noted that the operators defined in Eq.~\eqref{2.05} are not all Hermitian because the zeros of $S_{2M}$ may be complex, but the product operator $\ds\prod_{s=1}^{2M}K_s^{(2M)}$ is Hermitian.  We will see that a Hermitian version of the multipliers  permits further simplifications.

We note that $S_{2M}$ has  zeros which are either real or complex.  Furthermore, if $z_s$ is a zero so are $z_s^*$ and $-z_s$.  We write $z_s=X+iY$, with  $X$ and $Y$ real and consider the cases of $z_s$ real and $z_s$ complex separately. If $z_s$ is real, the set of zeros contains $z_s$ and $-z_s$, and we combine the two factors involving these zeros (with $H\Delta t/\hbar=\cH$)
\beq
\label{6.01}
\left(1-\dfrac{\cH}{z_s}\right)
\left(1+\dfrac{\cH}{z_s}\right) 
= 1 -  \dfrac{\cH^2}{|z_s|^2}.
\eeq      
If  $z_s$ has a nonzero imaginary part, there are four zeros with the same value of $|z_s|$ because $z_s^*$ and $-z_s$ are also zeros.  These can be used to write the product of the corresponding four binomials as a product of two trinomials,
\beq
\label{6.02}
\renewcommand{\arraystretch}{3}
\begin{array}{l}
\left(1-\dfrac{\cH}{z_s}\right)
\left(1-\dfrac{\cH}{z_s^*}\right)
\left(1+\dfrac{\cH}{z_s}\right)
\left(1+\dfrac{\cH}{z_s^*}\right)  \\
= \left(1+\dfrac{\cH^2}{|z_s|^2}-2X\dfrac{\cH}{|z_s|^2}\right)\left(1+\dfrac{\cH^2}{|z_s|^2}+2X\dfrac{\cH}{|z_s|^2}\right)
\end{array}
\eeq
When the operator $\cH^2$ is involved, $\cH$ acts on the wave function successively.  For example, when $2M=6$, there are six zeros, two real ones and four complex ones, leading to Eq.~\eqref{6.01} being applied once and Eq.~\eqref{6.02} once.  For convenience we write the product operator as 
\beq
\label{6.03}
B^{(2M,r)} = 2i\cH\prod_{\sigma=0}^{\sigma_{\max}}L^{(2M)}_\sigma,
\eeq
where each $L^{(2M)}_\sigma$ is a Hermitian operator of the form of Eqs.~\eqref{6.01} or \eqref{6.02} and $L^{(2M)}_0=1$.  We have recalculated the results of the example in Sec.~\ref{barrier} and obtained the same results with similar CPU times.
 
\subsection{Real and imaginary parts of the wave function at staggered times}
\label{visscher}

An approach that is efficient and accurate involves an explicit staggered-time algorithm.\cite{visscher91}  This second-order accurate-in-time method entails defining the real and imaginary parts of the wave function at alternate times.  Given that the operators $L_\sigma^{(M)}$ are real (Hermitian), the method can be generalized to yield solutions of higher-order accuracy in time and space.  Given the basic equation
\beq
\label{6.04}
\psi^{n+1} = \psi^{n-1} - 2i\cH\prod_{\sigma=0}^{\sigma_{\max}}L^{(2M)}_\sigma \psi^n,
\eeq
we write
\beq
\label{6.05}
\psi^{n} = \cR^n + i\cI^n, 
\eeq
where both $\cR^n$ and $\cI^n$ are real, and substitute for $\psi^{n}$ in Eq.~\eqref{6.04} to obtain
\begin{subequations}
\label{6.06}
\begin{align}
\cR^{2n} & = \ds\cR^{2n-2} +2\cH\prod_\sigma L_\sigma^{(2M)}\cI^{2n-1}  \\
\cI^{2n+1} & = \ds\cI^{2n-1} - 2\cH\prod_\sigma L_\sigma^{(2M)} \cR^{2n}
\end{align}
\end{subequations}
for $n=1,2,\dots,N$.   If  $\cR^0_j=\cR(x_j,0)$ and $\cI^1_j=\cI(x_j,\Delta t)$ for all $j$ are given as initial conditions, one generates $\cR(x_j,2n\Delta t)$ and $\cI(x_j,(2n+1)\Delta t)$ alternately.  Thus the real part of the wave function is calculated at even  time intervals and the imaginary part at odd  time intervals.

Following Visscher~\cite{visscher91} we define the probability density as
\begin{subequations}
\label{6.08}
\begin{align}
\rho^{2n}_j & = \left(\cR^{2n}_j\right)^2 + \cI^{2n+1}_j\cI^{2n-1}_j,  \\
\rho^{2n+1}_j & = \cR^{2n+2}_j\cR^{2n}_j + \left(\cI^{2n+1}_j\right)^2.
\end{align}
\end{subequations}
It is not difficult to show that the normalization using even numbers of time intervals is equal to that using odd numbers of time intervals.
An obvious computational advantage is that we deal with real numbers only.   The computed wave functions are accurate to order $(\Delta t)^{2M+1}$ in time and $(\Delta x)^{2r-1}$ in space.  Our formalism reduces to Visscher's algorithm\cite{visscher91} when $M=0$ and $r=1$.  Higher values of $M$ and $r$ significantly increase the precision and the efficiency of the computations.

This formulation is a significant extension of the Yee finite-difference time-domain approach, which was proposed for solving Maxwell's equations.\cite{yee66} More recently,   it was applied to solving the time-dependent \SE,\cite{nagel09,zhu21,napoles-duarte18} but tends to be used  with the forward Euler method Eq.~\eqref{1.05} rather than the three-time-point formula of Askar and Carmak in Eq.~\eqref{1.07}.  Typically the method is employed with $M=0$ and $r=1$.\cite{nagel09}   Some spatial high-order methods, i.e., with $r>1$, have been studied,\cite{zhu21,napoles-duarte18} but they still employ  a time-step-advance corresponding to $M=0$.

\subsection{Determination of the real and imaginary parts of the wave function independently}
\label{gray}
Gray and Balint-Kurti~\cite{gray98} describe an approach by which   the real part of the wave function is calculated without calculating the imaginary part.  This approach  is useful to generate the $S$-matrix elements used in scattering theory which are fully determined from the real part of the wave function.

By adding Eqs.~\eqref{1.04} and \eqref{1.06}, we obtain
\beq
\label{6.09}
\psi(\br,t+\Delta t) = -\psi(\br,t-\Delta t) +2\cos(\cH)\psi(\br,t),
\eeq
which is valid for the real and the imaginary parts of the wave function independently because $\cos(\cH)$ is a real operator.  The operator $\cos(\cH)$ can be expanded in a similar manner as $\sin(\cH)$.  We write
\beq
\label{6.10}
\cos(z) = C_{2\cL}(z) + \cO(z^{2\cL+2}),
\eeq
where the polynomial approximation of $\cos z$ is
\begin{subequations}
\label{6.11}
\begin{align}
C_{2\cL}(z) & = 1 - \dfrac{1}{2!}z^2 + \dfrac{1}{4!}z^4 - \cdots + \dfrac{(-1)^\cL}{(2\cL)!}z^{2\cL} \\
& = \ds\prod_{s=1}^{2\cL} \left(1-\dfrac{z}{z_s^{(2\cL)}}\right).
\end{align}
\end{subequations}
The first few zeros of the polynomial approximation $C_{2\cL}$ are given in Table~\ref{table_4}.

\renewcommand{\arraystretch}{1}
\begin{table*}[ht]
\caption{The zeros $z_s^{(2\cL)}$ of the polynomial approximation of $\cos(z)$ for $\cL=1,2,3$.}\label{table_4}
\footnotesize{
\begin{tabular}{c|cccccc}
\hline\hline
$\cL$ & $s=1$ & 2 & 3 & 4 & 5 & 6 \\
\hline
1 & -1.41421+\i0.00000 & 1.41421+\i0.00000 \\
2 & 1.59245+\i0.00000 & -1.59245+\i0.00000 & 3.07638+\i0.00000 & -3.07638+\i0.00000 \\
3 & ~~1.56991+\i0.00000~~ & ~~-1.56991+\i0.00000~~ & ~~3.92808-\i1.28924~~ & ~~-3.92808+\i1.28924~~ & ~~3.92808+\i1.28924~~ & ~~-3.92808-\i1.28924~~\\
\hline\hline
\end{tabular}
}
\end{table*}

We define the operator binomials
\beq
\label{6.12}
\cP_s^{(2\cL)} \equiv 1 - \dfrac{\cH}{z_s^{(2\cL)}}, \quad s = 1,2,\dots,2\cL,
\eeq
and obtain the time-advance expressions
\beq
\label{6.13}
\psi^{n+1} = -\psi^{n-1} + 2\prod_{s=1}^{2\cL} \cP_s^{(2\cL)}\psi^n.
\eeq
Because the operator $\cos{\cH}$ is real, we start with the real parts of $\psi^0$ and $\psi^1$ and generate the real part of $\psi^n$.   Similarly, we could begin with the imaginary parts and generate the imaginary part of the wave function at later times. If only the real part is needed, this method is  efficient.  We also note that Eqs.~\eqref{6.01} and \eqref{6.02} should be used   to avoid complex wave functions at intermediate stages.  Because this method involves real functions only at all stages, it simplifies the coding of the algorithm. We have replicated the results of Sec.~\ref{barrier} using this method.

\section{Discussion}\label{sec:disc}
We have reviewed an algorithmic approach of numerically solving the time-dependent Schr\"odinger equation on a spatial and temporal grid.  The main and novel focus is the expansion of the time-evolution as a function of the time increment.  This expansion is straightforward to implement to any order of the time increment and shows significant improvement of the accuracy even at high order. Among the various options of spatial integration we choose the general central differencing approach which yields wave functions accurate to a variety of orders of the spatial increment.  Thus a systematic expansion of the wave function of arbitrary order of accuracy in spatial and temporal increments is achieved. 

The method is an extension of the method of Askar and Cakmak\cite{askar78} used to obtain solutions to arbitrary order of $\Delta t$ and $\Delta x$. Their method is the $M=0$ and $r=1$ case of the method we have discussed.  Higher values of $M$ and $r$ yield solutions with significantly improved accuracy and efficiency. Even at larger values of $M$, e.g., $M>5$, the error can be reduced by orders of magnitude by increasing $M$.  Because of its  explicit nature, the method is more efficient than the implicit methods, especially for higher spatial dimensions.  Although normalization is not strictly conserved, the degree of normalization can be used as an indicator of the stability of the method.

An advantage of this approach  to the time-dependence is that once the algorithm is established, it can be applied to any order.  For instance, it does not suffer from the complication of expansions of the exponential  split-order methods which  are no longer practical for $\cO(\Delta t^{2M})$ with $M>3$.\cite{bandrauk92}  Furthermore, the method is easily   generalized to the staggered-time method of Visscher~\cite{visscher91} or the real part of the wave function calculation.\cite{gray98} 
 
\section{Suggested Problems}
\label{sec:problem}
\begin{enumerate}
\item Suppose that there is no closed form expression for the wave function. Write  pseudocode to determine the  wave function $\psi(x,\Delta t)$, where the input is $\psi(x,0)$ using the method outlined in Appendix~A for an arbitrary $r$ and $\cM$.
 
\item Consider the tunneling ionization of an atom in an electric field.  One-dimensional versions of this problem have been studied recently to elucidate features of the process.\cite{xu18,yusofsani20,gharibnejad20}  A model with $e=\hbar=m=1$, and the Coulomb gauge, employs a soft-core Coulomb potential to bind an electron
\beq
\label{p.01}
V_c(x) = -\dfrac{1}{\sqrt{1+x^2}},
\eeq
and a square-pulse laser field interaction
\beq
\label{p.02}
E(t) = E_0\theta(T-t),
\eeq
so that the effective interaction is given by the potential
\beq
\label{p.03}
V(x) = V_c(x) - E(t)x.
\eeq
\begin{enumerate}
\item Verify that 
\beq
\label{p.04}
\psi_\mr{gr}(x) = N(1 + \sqrt{x^2 +2})e^{\ts -\sqrt{x^2 + 2}}
\eeq
 is the ground state wave function of the soft Coulomb potential with energy $-\dfrac{1}{2}$.
 
\item Plot $V(x)$ for $E_0=0.05$ and determine the classical turning points at the Coulomb ground-state energy.

\item Write a program to solve the time-dependent Schr\"odinger equation with $V(x)$ given by Eq.~\eqref{p.03} with   $E_0=0.1$ and the initial wave function $\psi(x,0) = \psi_\mr{gr}(x)$.  Use the method of Problem 1 to determine $\psi(x,\Delta t)$ and then the iterative method involving the expansion of $\sin\left(\dfrac{\Delta t H}{\hbar}\right)$ to obtain the wave functions on a time grid.  To start set $x_{\min} = -200$, $x_{\max} = 1000$ with $J= 15000$, $t_{\max}=100$, $N=10,000$, $T=t_{\max}$, $\Delta t=0.01$, $r=1$, and $M=5$.

\item Experiment with smaller $E_0$, e.g., $E_0=0.05$ when the initial energy is below the barrier height.  One would expect a smaller ionization rate;  identify tunneling and atomic excitations followed by escape.
\item (More advanced.) The analysis of tunneling ionization often involves quantum tunneling followed by the classical motion of the ejected electron.  To obtain a classical trajectory of the ejected electron, we need the position at which the electron is ejected and an initial velocity.  One might consider the classical turning point as such a point at which the velocity is zero.  However, there is uncertainty as to whether experimental data fit such a prescription.\cite{xu18,yusofsani20}  Discuss in terms of the results of calculations.

\end{enumerate}
\end{enumerate}


\appendix

\section{Calculating $\psi^1$ from $\psi^0$}\label{app:01}

We require $\psi(x,0)$ and $\psi(x,\Delta t)$ as inputs to  the main algorithm of this paper.  Unless an exact solution is available, we need a method of determining $\psi(x,\Delta t)$ from $\psi(x,0)$ accurately.
We formulate the initial value problem using Eq.~\eqref{1.04}
\beq
\label{a.01}
\psi^{n+1} = e^{\ts -iH\Delta t/\hbar} \psi^n  \mbox{ for a given $\psi^0$}. 
\eeq
To determine $\psi^1$ accurately from $\psi^0$, we approximate the exponential function as
\beq
\label{a.02}
e^{\ts z} = \prod_{s=1}^\cM\left(1-\dfrac{z}{z^{(\cM)}_s}\right) + \cO(z^{\cM+1}),
\eeq 
where the $z^{(\cM)}_1,z^{(\cM)}_2,\dots,z^{(\cM)}_\cM$ are the $\cM$ zeros of the $\cM$th order 
polynomial approximation of the exponential function. These can be calculated to the precision needed, a sample of which for $\cM=1$ to 5    is given in Table~\ref{table:01} to five decimal places.

\begin{table*}[t]
\caption{The zeros $z_s^{(\cM)}$ of the polynomial approximation of $e^{\ts z}$ for $\cM$ from 1 to 5.}\label{table:01}
\renewcommand{\arraystretch}{1.3}
\begin{tabular}{cccccc}
\hline\hline
$\cM$ & $s=1$ & 2 & 3 & 4 & 5 \\
\hline
1 & $-1.00000+i0.00000$ \\
2 & -1.00000+\i1.00000 & -1.00000-\i1.00000 \\
3 & -1.59607+\i0.00000 & -0.70196-\i1.80734 & -0.70196+\i1.80734 \\
4 & -0.27056+\i2.50478 & -1.72944+\i0.88897 & -1.72944-\i0.88897 & -0.27056+\i2.50478 \\
5 & ~~+0.23981+\i3.12834~~ & ~~-1.64950+\i1.69393~~ & ~~-2.18061+\i0.00000~~ & ~~-1.64950-\i1.69393~~ & ~~+0.23981-\i3.12834~~ \\
\hline\hline
\end{tabular}
\end{table*}

The exponential function expansion can be employed to express the time-evolution operator as a product of binomial factors.  Define the operator
\beq
\label{a.03}
G_s^{(\cM)} \equiv 1 + \dfrac{iH\Delta t/\hbar}{z_s^{(\cM)}},
\eeq
so that
\beq
\label{a.04}
e^{\ts -iH\Delta t/\hbar} = \prod_{s=1}^\cM G_s^{(\cM)} +\cO\left[(\Delta t)^{\cM+1}\right].
\eeq
We use the exponential expansion in Eq.~\eqref{a.01} to write 
\beq
\label{a.05}
\psi^{n+1} = \prod_{s=1}^\cM G_s^{(\cM)}\psi^n.
\eeq
Because the expression involves the product of $\cM$ operators, we define intermediate functions $\psi^{(n+s/\cM)} = G_s^{(\cM)}\psi^{(n+(s-1)/\cM)}$, so that we solve for $\psi^{n+1}$ recursively, starting with 
\beq
\label{a.06}
\psi^{(n+1/\cM)} = G_1^{(\cM)}\psi^{(n)}.
\eeq
Once $\psi^{(n+1/\cM)}$ is determined, the process is repeated, to obtain in succession $\psi^{(n+2/\cM)}$, $\psi^{(n+3/\cM)}$, $\dots$, $\psi^{(n+(\cM-1)/\cM)}$, $\psi^{n+1}$.  At each step an equation like  Eq.~\eqref{1.05} is solved, the only difference is that effectively the real $\Delta t$ is replaced by a complex one.  Because the operators $G_s^{(\cM)}$ commute, they may be applied in any order.  To make the accuracy compatible with that of the expansion of the sine operator, we set $\cM=2M$.

Although the magnitude of the growth factor for $\cM=0,r=1$ is greater than one,\cite{press_F92} it can be less than one for larger values of $\cM$, and stable algorithms are possible.  For a single time step, stability is not an issue, and in any case we can check the precision by examining the normalization of $\psi^1$.

\section{Product of banded matrices}
\label{app:02}

Suppose that $A$ is a banded matrix with lower bandwidth $p$ and upper bandwidth $q$ (total bandwidth $p+q+1$) and $B$ is a  banded matrix with lower bandwidth $r$ and upper bandwidth $s$. Then the product matrix $C=AB$ is a banded matrix with lower bandwidth $p+r$ and upper bandwidth $q+s$.\footnote{For a proof see https://math.stackexchange.com/questions/1922862/product-of-banded-matrices.}

Consider  $\ds\prod_{s=0}^{2M} K_s^{(2M)}$ with $K_0^{(2M)} = I$, and $K_s^{(2M)}$ for $s\ge 1$ having lower and upper bandwidths of $r$ (total bandwidth $2r+1$). We denote the lower and upper bandwidths as $lb$ and $ub$, respectively, and thus
$K_2^{(2M)}K_1^{(2M)}$ has $lb=ub=2r$, 
$K_3^{(2M)}K_2^{(2M)}K_1^{(2M)}$  has $lb=ub=3r$, and 
$K_{2M}^{(2M)}K_{2M-1}^{(2M)}\cdots K_1^{(2M)}K_0^{(2M)}$ has $lb=ub=2Mr$.
We define
\beq
A= 2iH\Delta t/\hbar\prod_{s=0}^{2M}K_s^{(2M)},
\eeq
and note that $A$ has is a banded matrix with lower and upper bandwidths of $(2M+1)r$ or total bandwidth of $(4M+2)r+1$.

Note the multiplication of  banded matrix with lower and upper bandwidth $r$ with a banded matrix with lower and upper bandwidth $sr$ yields a banded matrix with lower and upper bandwidth $(s+1)r$.  In other words, $C^{([s+1]r)} = A^{(r)}B^{(sr)}$, or
\beq
\label{999}
c^{([s+1]r)}_{ij} = \sum_{k'=i-r}^{i+r} a^{(r)}_{ik'}b^{(sr)}_{k'j}
= \sum_{k=-r}^r a^{(r)}_{i,i+k}b^{(sr)}_{i+k,j},
\eeq
where the superscript is the lower and upper bandwidth.
Outside the banded diagonals the matrix elements are all zero.


\bibliography{apj_wvd.bbl}
       
\end{document}